\documentclass[a4paper]{article}
\usepackage{jheppub}
\usepackage{dsfont}
\usepackage{amsmath,amssymb,amscd,amsfonts,mathtools}

\usepackage[toc,page]{appendix}
\usepackage{epsfig}
\usepackage{epstopdf}
\usepackage{latexsym}
\usepackage{graphicx}
\usepackage{placeins}
\usepackage{floatrow}
\usepackage{subfig}
\usepackage{caption}
\usepackage{booktabs}
\usepackage{bbm}
\usepackage{color}
\usepackage{physics}
\usepackage{tensor}
\usepackage{tikz}
\usepackage{comment}
\usetikzlibrary{matrix}
\usetikzlibrary{decorations.markings,calc,shapes,decorations.pathmorphing,patterns,decorations.pathreplacing,arrows.meta}
\usetikzlibrary{positioning}
\usepackage{hyperref}

\pdfoutput=1
\makeatletter
\def\@fpheader{\relax}
\makeatother

\def\J{{\cal J}}

\newcommand\numberthis{\addtocounter{equation}{1}\tag{\theequation}}

\title{Background ambiguity and the G\"odel double copy}
\author{Brian Kent$^A$}
\author{Tucker Manton$^{B,C}$}
\author{Sanjit Shashi$^{A,D}$}
\affiliation{$^A$Theory Group, Weinberg Institute, Department of Physics, University of Texas,\\
\phantom{}\hspace{0.5cm} 2515 Speedway, Austin, Texas 78712, USA.}
\affiliation{$^B$Department of Physics, Brown University,\\
\phantom{}\hspace{0.5cm} 182 Hope Street, Providence, Rhode Island 02912, USA.}
\affiliation{$^C$School of Fundamental Physics and Mathematical Sciences, Hangzhou Institute for Advanced Study,\\
\phantom{}\hspace{0.5cm} University of Chinese Academy of Sciences (HIAS-UCAS), Hangzhou, 310024, China.}
\affiliation{$^D$Santa Cruz Institute for Particle Physics, Department of Physics, University of California,\\
\phantom{}\hspace{0.5cm} 1156 High Street, Santa Cruz, California 95064, USA.}
\emailAdd{brian\_kent@utexas.edu}
\emailAdd{tucker\_manton@ucas.ac.cn}
\emailAdd{sashashi@ucsc.edu}

\abstract{In this work, we investigate the assumptions regarding spacetime backgrounds underlying the classical double copy. We argue (contrary to the norm) that single-copy fields naturally constructed on the original curved background metric are only interpretable on a flat metric when such a well-defined limit exists, for which Kerr--Schild coordinates offer a natural choice. As an explicit example where such a distinction matters, we initiate an exploration of single-copies for the G\"odel universe. This metric lacks a (geodesic) Kerr--Schild representation yet is Petrov type-D, meaning the technology of the ``Weyl double copy" may be utilized. The Weyl derived single copy has many desirable features, including matching the defining properties of the spacetime, and being sourced by the mixed Ricci tensor just as Kerr--Schild single copies are. To compare, we propose a sourced flat-space single-copy interpretation for the G\"odel metric by leveraging its symmetries, and find that this proposal lacks the defining properties of the spacetime, and is not consistent with the flat limit of our curved-space single copy. Notably, this inconsistency does not occur in Kerr--Schild metrics. Our curved-space single copy also lead to the same electromagnetic analogue of the G\"odel universe found separately through tidal force analogies, opening a new avenue of exploration between the double copy and gravitoelectromagnetism programs.}

\begin{document}	
\maketitle
\flushbottom
\vfill\pagebreak

\section{Introduction}

%Structure of equations hidden within each other
%Start with amplitudes
%Start with 2 different nonequivalent double copies
%Table of vacuum & sourced vs Weyl and KS

The term \textit{double copy} finds its roots in the amplitudes program, 
where it was discovered by Bern, Carrasco, and Johansson (BCJ) that graviton scattering amplitudes can be computed from nonabelian gauge theory amplitudes by a simple process entailing exchanging color factors (contractions of structure constants) with kinematic numerators (functions of momenta and polarizations) \cite{Bern:2008qj, Bern:2010ue}. This is known as BCJ or color-kinematics duality and has proven to be extremely useful across a surprising range of applications (see \cite{Bern:2019prr} for a thorough review). At tree level, double copy relations are equivalent to the Kawai--Lewellen--Tye (KLT) relations, which relate open string amplitudes to closed string amplitudes \cite{Kawai:1985xq}. However the double copy is conjectured to hold at all loop orders and has been employed in calculations up to four loops \cite{Bern:2012uf}. This hints towards the possibility of a nonperturbative (with respect to Newton's constant) double copy relation. Such a relation is known to exist at the level of exact solutions and is referred to as the \textit{classical double copy}, which can be thought of as a modern avatar of some striking structural correspondences between general relativity and classical electromagnetism.

%where it was revealed through BCJ/color-kinematics duality \cite{Bern:2008qj, Bern:2010ue} that graviton scattering amplitudes could be obtained from gauge theory amplitudes through the exchange of color and kinematic information. This relation at tree-level could be equivalently gleaned from the low energy limit of the KLT relations \cite{Kawai:1985xq}, which relate closed and open string scattering amplitudes. An excellent review of this correspondence can found in \cite{Bern:2019prr}, while its connection to the present subject of interest, \textit{the classical double copy}, can be readily found in most introductions of papers on the subject. Of concern to our present work, the classical double copy can be described as a modern avatar of some striking structural correspondences between general relativity and classical electromagnetism. 

Connections between these two fields is as old as general relativity itself, but this current program of research investigates how several physically relevant spacetimes can be constructed from ``squaring" some quantity in the gauge theory to obtain an analogous quantity in the gravity theory. %(which is practically investigated in the opposite manner).
This can be in the form of squaring an electromagnetic potential equivalent to a metric perturbation of Einstein's equations (the \textit{Kerr--Schild double copy}) \cite{Monteiro:2014cda}, or squaring an electromagnetic spinor to realize a Weyl spinor capturing the gravitational curvature
%how the pure gravitational degrees of freedom within the curvature, captured by the Weyl spinor, may be decomposed into a squared electromagnetic spinor
(the \textit{Weyl double copy}) \cite{Luna:2018dpt}. Both of these constructions have been explored extensively in many different settings, such as nonsingular spacetimes \cite{Mkrtchyan:2022ulc,Easson:2020esh}, higher dimensional spacetimes \cite{Zhao:2024ljb,Ortaggio:2023cdz,Chawla:2022ogv}, gravitational waves \cite{Godazgar:2020zbv,CarrilloGonzalez:2022mxx,Andrzejewski:2019hub,Ilderton:2018lsf}, and gravitational horizons \cite{Chawla:2023bsu,He:2023iew} to name a few.

%\cite{Luna:2015paa, White:2016jzc, Goldberger:2016iau, DeSmet:2017rve, Adamo:2017nia, Easson:2023dbk, Lee:2018gxc, Gurses:2018ckx, Bahjat-Abbas:2018vgo, Bah:2019sda, Cho:2019ype, Alawadhi:2019urr, Banerjee:2019saj, Kim:2019jwm, Huang:2019cja, Arkani-Hamed:2019ymq, CarrilloGonzalez:2019gof, Luna:2020adi, Keeler:2020rcv, Easson:2020esh, Elor:2020nqe, prabhu2020classical, alawadhi2020s, cheung2020scattering, delaCruz:2020bbn, Emond:2020lwi, White:2020sfn, 1969JMP....10.1842D, Godazgar:2020zbv, Chacon:2020fmr, Berman:2020xvs, Ben-Shahar:2021zww, Monteiro:2021ztt, guevara2021reconstructing, Gonzalez:2021bes, Chacon:2021hfe, Cho:2021nim, Chacon:2021wbr, Easson:2021asd, Campiglia:2021srh, Farnsworth:2021wvs, guevara2021worldsheet, Adamo:2021dfg, Godazgar:2021iae, Kosower:2022yvp, CarrilloGonzalez:2022mxx, Alkac:2021seh, Alkac:2021bav, Chacon:2021lox, Chen:2021chy, Gonzalez:2021ztm, Angus:2021zhy, Moynihan:2021rwh, Alkac:2022tvc, Han:2022ubu, Mao:2021kxq, Nagy:2022xxs, Armstrong-Williams:2022apo, CarrilloGonzalez:2022ggn, Ben-Shahar:2022ixa, Didenko:2022qxq, Easson:2022zoh, Chawla:2022ogv, Armstrong-Williams:2023ssz, chawla2023black, lipstein2023self, gonzalez2023mini, Bonezzi:2023pox, Brown:2023zxm, Borsten:2023paw, Ball:2023xnr}.
%\sanjit{we don't need to cite all of these. imo we should just pick the ones that actually tread new ground into the ``many different contexts" rather than citing every smaller thing that has been done; otherwise the list is not actually useful to a reader.}

Explorations of why these three seemingly unrelated mathematical frameworks---the amplitude, Kerr--Schild, and Weyl double copies---are related have taken many routes. Much progress has been taken in expanding and developing the double copy relation for Kerr--Schild metrics; this includes utilizing the Newman--Penrose formalism \cite{Elor:2020nqe, Farnsworth:2021wvs}, and allowing an extension of the Kerr--Schild double copy to type II spacetimes \cite{Easson:2023dbk}. Connections between the momentum space (amplitudes) double copy and classical double copy have been investigated through the language of twistor theory \cite{Chacon:2021wbr,Chacon:2021lox,Luna:2022dxo}, leading to the Weyl double copy in position space. The fluid-gravity correspondence offers an interesting landscape with which to study the interplay between background geometries and perturbations \cite{Keeler:2020rcv,Keeler:2024bdt}. Furthermore, the Weyl double copy has also been used to study sourced spacetimes \cite{Easson:2021asd,Easson:2022zoh,Armstrong-Williams:2024bog,Armstrong-Williams:2023ssz,Alkac:2023glx}, which will be relevant in our discussion of the G\"odel solution. These directions are just a small sample within a plethora of rich literature.

However when considering exact solutions of Einstein's equations at nonlinear level, it is still generally unknown what assumptions precisely underlie this correspondence, which is the subject of this current work. One crucial step in this direction lies in an ambiguity still present within both the Kerr--Schild and Weyl double copies---the geometry of the background on which the ``single-copy" fields are constructed. We should note that this is a rather different investigation to \cite{Carrillo-Gonzalez:2017iyj}, which considers Kerr--Schild perturbations to nonflat base metrics. Rather, our investigation concerns the ambiguity often left undisclosed; both the Kerr--Schild and Weyl double copies have valid single-copy interpretations on either the original curved background or an ``appropriate" flat background. Furthermore, the form of the electromagnetic field strength and source is equivalent in both prescriptions. That this ambiguity exists is a testament to the power of our coordinate choice; Kerr--Schild is in a sense an ``exact" linearization of the metric, such that higher-order terms (with respect to the Kerr--Schild perturbation) of the curved space Maxwell equations contract to zero. 

Hence it is really the case that the flat space interpretation in Kerr--Schild coordinates emerges from the full curved spacetime. This has two prominently desirable features: a flat background removes any effects of gravity therefore isolating the electromagnetic effects, and connection with the motivating case of amplitudes may be more directly considered, where the amplitudes themselves are only well understood on a flat background. However when seeking single-copies beyond the Kerr--Schild double copy from the structure of GR itself (as suggested by the Weyl double copy), the dependence on a particular choice of coordinates should not be fundamental.

We explicitly mention that searching for a single copy is \textit{not} the same as creating a consistent electromagnetic field sourcing our curved spacetime, which would properly account for the interplay of electromagnetism and gravity. Rather, the classical double copy suggests a procedure to convert curvature quantities and Einstein's equations into objects that look like electromagnetic fields satisfying Maxwell's equations. That the former is only nontrivial on a curved background suggests the latter should be constructed on that background. There is however still great utility in the ability to take a flat limit, but it must be carried out carefully.

To provide a specific example where such a distinction is pushed to the forefront, we examine a specific manifestation of the type-D Weyl double copy that is beyond the flat-space Kerr--Schild double copy. We focus on the G\"odel metric, a homogeneous, one-parameter, and sourced spacetime that is most famous for its inclusion of naked closed timelike curves \cite{Godel:1949ga}. The G\"odel metric can be smoothly deformed to Minkowski space. However, we find that, since G\"odel cannot be written in geodesic Kerr--Schild form, we lose the special property that the electromagnetic field and source are equivalent in this limit. By choosing the Weyl double copy as our starting point, we take the stance that it is the spinor formalism of gravity that most naturally represents the classical double copy, a feature broadly shared for many structural questions in general relativity \cite{Penrose_Rindler_1984}.

We stress at this point that our G\"odel single copy can at worst be taken as a guess, since no well-defined non-Kerr--Schild sourced classical double copy prescription has been proven to exist in generality thus far. However this guess has several desirable features suggesting that the original curved background is naturally where a single copy of the G\"odel spacetime should be constructed. Without introducing new arbitrary functions of the coordinates within the electromagnetic field (which one can force to satisfy Maxwell's equations by defining an appropriate source)---a procedure that would be ill-defined without a proper formalism---our single copy, along with its scaled duality rotations, is the only single copy consistent with the Weyl double copy and furthermore has a non-zero source on its original curved background. Choosing a physical solution with no magnetic-charges yields a fantastic correspondence: although the G\"odel spacetime is not of Kerr--Schild form, the correspondence of sources between Maxwell and Einstein solutions for Kerr--Schild spacetimes, $J^\mu \sim \tensor{R}{^\mu_0}$, appears to still hold. Conversely, if one tries to take the Kerr--Schild prescription sources as a starting point, one is led to our single copy via the symmetries of the spacetime. Furthermore our single copy matches the prominent features of the G\"odel spacetime: symmetries, homogeneity, and it being ``everywhere rotating." We compare this procedure to enforcing a sourced single copy on flat space by leveraging the symmetries of the spacetime. We find in this case not only a far more complicated single copy with many unconstrained variables, but also one in which almost none of the primary characteristics of the G\"odel spacetime have a natural interpretation.

We proceed as follows. In Section \ref{sec:reviewSpin}, we present the necessary mathematical background to understand the Weyl double copy through the language of spinors, emphasizing several desirable benefits from using such a formalism. Proceeding this, in Section \ref{sec:structureCDC} we lay out the structures and correspondence between the two primary formulations of the classical double copy, clearing up some misconceptions present in the literature, and providing a well-defined procedure clarifying how one may interpolate from the originally curved space to a flat one. In Section \ref{sec:godel}, we present the G\"odel metric, the spacetime of our current investigations along with its key properties that should be present in an electromagnetic single copy. Finally, in Section \ref{sec:SingleCopy} we exhaust the reasonable electromagnetic single-copies of G\"odel and make an explicit single-copy correspondence.

\section{Reviewing spinors in general relativity}\label{sec:reviewSpin}

The Weyl double copy makes use of the spinorial formulation of gravity \cite{Penrose_Rindler_1984}, which in turn has direct links to the Newman--Penrose (NP) formalism \cite{Witten:1959zza,Penrose:1960eq,Newman:1961qr}, where one formulates 4d general relativity in the language of a complex null tetrad. See \cite{Wald:1984rg,Stephani:2003tm,ODonnell:2003lqh} for extensive reviews of the subject. In hopes of linking those more familiar with the NP formalism to the spinor formalism, we present the necessary mathematical background within the language of both. Wherever necessary later in the manuscript, we freely call upon the results reviewed in this section.

\subsection{Basics of the tetrad formalism}

Standard general relativity chooses the tangent-space basis to be the coordinate basis. However, when a Minkowski basis (or any other noncoordinate basis) is chosen, we arrive at the \textit{tetrad formulation} of gravity. These bases need not be related by coordinate transformations, only invertible matrices to ensure the new basis spans the space. 

At every point, we may consider an orthonormal Minkowski basis $\{\mathbf{e}^a\}$ ($-+++$) such that the metric on the space is $\eta_{ab}$ %(using Latin lowercase indices for tetrad indices)
and the indices $a = 0,1,2,3$ are a four-fold ``flat" index. To transform our Minkowski basis to the standard coordinate basis $dx^\mu$, we employ the matrix $e_a^\mu$, called the (inverse) \textit{vierbein}, that maps our Minkowski basis to the coordinate basis via $dx^\mu = e_a^\mu \mathbf{e}^a $. The vierbein $e^a_\mu$ may be utilized to arrive back at the coordinate basis metric:
\begin{equation}
g_{\mu\nu} = e^a_\mu e^b_\nu \eta_{ab},\ \ \ \ e^a_\mu e^\mu_b = \delta^a_b.
\end{equation}
Basically, $e_\mu^a$ allows us to switch between the flat indices and the Greek spacetime (``curved") indices. As an alternative, one may take a complex null tetrad as the basis. The four vectors are
\begin{equation}
\ell^a \equiv \frac{1}{\sqrt{2}}(\delta_0^a + \delta_3^a),\ \ \ \ 
n^a \equiv \frac{1}{\sqrt{2}}(\delta_0^a - \delta_3^a),\ \ \ \ 
m^a \equiv \frac{1}{\sqrt{2}}(\delta_1^a - i\delta_2^a),\ \ \ \ 
\overline{m}^a = \frac{1}{\sqrt{2}}(\delta_1^a + i\delta_2^a).\label{nulltetfromOrtho}
\end{equation}
The above vectors are null and also satisfy the following relations:
\begin{equation}
l^an_a = -1, \qquad m^a \overline{m}_a = 1, \qquad \eta_{ab} = 2m_{(a}\overline{m}_{b)} - 2\ell_{(a}n_{b)}.\label{cplxnullgeneric}
\end{equation}
Utilizing the vierbein yields the curved index version of these vectors:
\begin{equation}
\ell^\mu \equiv \frac{1}{\sqrt{2}}(e_0^\mu + e_3^\mu),\ \ \ \ 
n^\mu \equiv \frac{1}{\sqrt{2}}(e_0^\mu - e_3^\mu),\ \ \ \ 
m^\mu = \frac{1}{\sqrt{2}}(e_1^\mu - ie_2^\mu),\ \ \ \ 
\overline{m}^\mu = \frac{1}{\sqrt{2}}(e_1^\mu + ie_2^\mu).\label{nulltetfromOrthocurved}
\end{equation}
One key result of choosing the complex null tetrad is that we can fully characterize the algebraic structure of the Weyl tensor $W$ in terms of five different contractions of $W$, collectively called the \textit{Weyl scalars}:
\begin{equation}
\begin{Bmatrix*}[l]
\Psi_0 \equiv W_{\mu\nu\lambda\rho}\ell^\mu m^\nu \ell^\lambda m^\rho,\\
\Psi_1 \equiv W_{\mu\nu\lambda\rho}\ell^\mu m^\nu \ell^\lambda n^\rho,\\
\Psi_2 \equiv W_{\mu\nu\lambda\rho}\ell^\mu m^\nu \overline{m}^\lambda n^\rho,\\
\Psi_3 \equiv W_{\mu\nu\lambda\rho}\ell^\mu n^\nu \overline{m}^\lambda n^\rho,\\
\Psi_4 \equiv W_{\mu\nu\lambda\rho}\overline{m}^\mu n^\nu \overline{m}^\lambda n^\rho.
\end{Bmatrix*}\label{weylScalarsDef}
\end{equation}
These scalars have a simpler, almost trivial interpretation upon the introduction of spinors, as we will soon see.

\subsection{Going to spinorland}
A powerful structural tool available via the tetrad formalism is to reformulate GR in the language of spinors, and its power lies in the simple structure assumed by tensorial quantities when recast in this manner. That we can use spinors in spacetime is motivated from the tetrad formalism: at any given point, two Minkowski observers (tetrad bases) are related by Lorentz transformations, and by restricting to the proper orthochronous subgroup, we utilize $SO^+(1,3) \cong SL(2,\mathbb{C})$. Therefore any Minkowski basis (and associated transformations) can be represented through ``left-handed" (chiral) and ``right-handed" (anti-chiral) spinor indices.

\begin{figure}
\centering
\begin{tikzpicture}[scale=0.7]
\node (A) at (0,3.6) {$\{A,B,\dots,A',B',\dots\}$};
\node (B) at (2,0) {$\{\mu,\nu,\dots\}$};
\node (C) at (-2,0) {$\{a,b,\dots\}$};

\draw[<->,thick] (A) to node [right] {$\sigma_\mu^{AA'}$} (B);
\draw[<->,thick] (A) to node [left] {$\sigma_a^{AA'}$} (C);
\draw[<->,thick] (C) to node [below] {$e_\mu^a$} (B);
\end{tikzpicture}
\caption{The different types of indices: curved (Greek), flat (lowercase Latin), and spinor (uppercase Latin, with unprimed indices describing chiral degrees of freedom and primed indices representing anti-chiral degrees of freedom). The mixed-index objects that translate between these different index types are also shown.}
\label{figs:indexMath}
\end{figure}
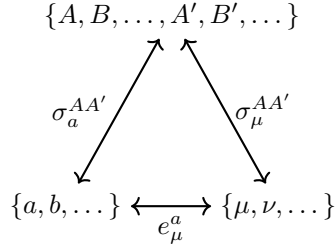

We label chiral spinor degrees of freedom with unprimed capital Latin indices, e.g. $\psi^A$ where $A \in \{+,-\}$, whereas we denote anti-chiral spinor degrees of freedom with primed capital Latin indices, e.g. $\overline{\psi}^{A'}$ with $A' \in \{+',-'\}$. Focusing on the chiral sector, the spinor indices are raised and lowered by a two-index Levi-Civita symbol [by virtue of being the invariant $SL(2,\mathbb{C})$ tensor under the Lorentz group]:
\begin{equation}
\psi_A = \psi^B \epsilon_{BA},\ \ \psi^{A} = \epsilon^{AB}\psi_{B},
\end{equation}
with a similar rule for anti-chiral indices. It is useful to define a dyad of basis chiral spinors (and their complex conjugates, which form a basis for the anti-chiral sector). We write these as $\{o^A,\iota^A\}$, and they are defined by the relations
\begin{equation}
o_A \iota^A = -o^A \iota_A = 1,\ \ o_A o^A = \iota_A \iota^A = 0.
\end{equation}
The lowered and raised Levi-Civita symbols in terms of these basis spinors are respectively
\begin{equation}
\epsilon_{AB} = o_A \iota_B - \iota_A o_B,\ \ \epsilon^{AB} = o^A \iota^B - \iota^A o^B.
\end{equation}
To switch between spinor and flat indices, we define an object called the ``Pauli vector" $\sigma_a^{AA'}$. First, we write the identity and the three Pauli matrices (up to an overall normalization of $\frac{1}{\sqrt{2}}$):
\begin{equation}
\begin{matrix*}[l]
\sigma^{AA'}_0 = \dfrac{1}{\sqrt{2}} (o^A \overline{o}^{A'} + \iota^A \overline{\iota}^{A'}), &&& \sigma^{AA'}_1 = \dfrac{1}{\sqrt{2}}(o^A \overline{\iota}^{A'} + \iota^A \overline{o}^{A'}),\vspace{0.2cm}\\
\sigma^{AA'}_2 = \dfrac{i}{\sqrt{2}}(o^A \overline{\iota}^{A'} - \iota^A \overline{o}^{A'}), &&& \sigma^{AA'}_3 = \dfrac{1}{\sqrt{2}}(o^A \overline{o}^{A'} - \iota^A \overline{\iota}^{A'}).
\end{matrix*}
\end{equation}
The Pauli vector is then
\begin{equation}
\sigma_a^{AA'} = \sigma^{AA'}_{0} \delta_a^0 + \sigma_1^{AA'} \delta_a^1 + \sigma_2^{AA'} \delta_a^2 + \sigma_3^{AA'} \delta_a^3,\label{paulivec}
\end{equation}
and we can compute the following contractions:
\begin{equation}
(\sigma_a^{AA'}\sigma_b^{BB'} + \sigma_b^{AA'}\sigma_a^{BB'})\epsilon_{A'B'} = -\eta_{ab}\epsilon^{AB},\ \ \ \ \sigma_{AA'}^a \sigma_b^{AA'} = -\delta^a_b,\ \ \ \ \sigma_{AA'}^a \sigma_a^{BB'} = -\delta_A^B \delta_{A'}^{B'}.
\end{equation}
We can also define the contraction $\sigma_{\mu}^{AA'} \equiv e_\mu^a \sigma_a^{AA'}$. This switches between spinor indices and curved indices. The different rules of how to switch between the index types are summarized in Figure \ref{figs:indexMath}.

\paragraph{Constructing the Weyl spinor} In the language of spinors, the elements of the complex null tetrad \eqref{nulltetfromOrtho} take on simple forms. We write them as follows:
\begin{align*}
\ell^{\mu} \sigma_\mu^{AA'} &= \frac{1}{\sqrt{2}}(\sigma_0^{AA'} + \sigma_3^{AA'}) = o^A \overline{o}^{A'},&
n^\mu \sigma_\mu^{AA'} &= \frac{1}{\sqrt{2}}(\sigma_0^{AA'} - \sigma_3^{AA'}) = \iota^A \overline{\iota}^{A'},\\
m^\mu \sigma_\mu^{AA'} &= \frac{1}{\sqrt{2}}(\sigma_1^{AA'} - i\sigma_2^{AA'}) = o^A \overline{\iota}^{A'},&
\overline{m}^{\mu}\sigma_\mu^{AA'} &= \frac{1}{\sqrt{2}}(\sigma_1^{AA'} + i\sigma_2^{AA'}) = \iota^A \overline{o}^{A'}. \numberthis
\end{align*}
%which is why the NP tetrad may also be called the ``induced" tetrad from the spinor basis.
Furthermore, we can exploit the symmetries of the Weyl tensor to write it as follows:
\begin{equation}
W_{\mu\nu\lambda\rho}\sigma^\mu_{AA'}\sigma^\nu_{BB'}\sigma^{\lambda}_{CC'}\sigma^\rho_{DD'} = \Psi_{ABCD}\epsilon_{A'B'}\epsilon_{C'D'} + \overline{\Omega}_{A'B'C'D'}\epsilon_{AB}\epsilon_{CD}.
\end{equation}
$\Psi$ is the chiral Weyl spinor while $\overline{\Omega}$ is the anti-chiral Weyl spinor. Both have totally symmetric index structures. Notably, if the Weyl tensor is real, then the entire combination must be invariant under complex conjugation, and this implies
\begin{equation}
(\Psi_{ABCD})^* = \overline{\Omega}_{A'B'C'D'} \equiv \overline{\Psi}_{A'B'C'D'}.
\end{equation}
Our focus here will be on metrics with real Weyl tensors, so we hereafter refer to $\Psi$ as \textit{the} Weyl spinor (which itself is generically complex). By expanding in terms of a basis, the totally symmetric Weyl spinor has a very simple decomposition into outer products of basis spinors, with each independent coefficient being one of the five Weyl scalars defined in \eqref{weylScalarsDef}:
\begin{equation}
\begin{split}
\Psi_{ABCD} =\ \ &\Psi_{0}\,\iota_A \iota_B \iota_C \iota_D -4\Psi_1\,\iota_{(A}\iota_B \iota_C o_{D)} + 6\Psi_2\,\iota_{(A}\iota_B o_C o_{D)}\\
&- 4\Psi_3\,\iota_{(A}o_B o_C o_{D)} + \Psi_{4}\,o_A o_B o_C o_D.
\end{split}\label{basisweyl}
\end{equation}
This can be shown by rewriting the Weyl scalars as contractions of chiral spinors:
\begin{equation}
\begin{Bmatrix*}[l]
\Psi_0 = \Psi_{ABCD}\, o^A o^B o^C o^D,\\
\Psi_1 = \Psi_{ABCD}\, o^A o^B o^C \iota^D,\\
\Psi_2 = \Psi_{ABCD}\, o^A o^B \iota^C \iota^D,\\
\Psi_3 = \Psi_{ABCD}\, o^A \iota^B \iota^C \iota^D,\\
\Psi_4 = \Psi_{ABCD}\, \iota^A \iota^B \iota^C \iota^D.
\end{Bmatrix*}
\label{weylScalarsDefSpinorland}
\end{equation}

\paragraph{The Petrov classification}

The Petrov classification organizes spacetimes based on the uniqueness of the (up to) 4 independent eigenbivectors of the Weyl tensor. In spinorial form, this classification becomes remarkably simple. However, to understand the classification in spinor language we must first understand the properties of totally symmetric spinors, like the Weyl spinor $\Psi_{ABCD}$. 

Any totally symmetric spinor $K_{A_1 ... A_n} = K_{(A_1...A_n)}$ can be rewritten as a symmetrized outer product of 1-spinors \cite{Penrose_Rindler_1984}:
\begin{equation}
 K_{A_1...A_n} = \xi^{(1)}_{(A_1}\xi^{(2)}_{A_2}...\xi^{(n)}_{A_n)} \label{symmspinor}
\end{equation}
Where each $\xi^{(i)}_A$ is a \textit{principal null spinor}. When this decomposition is applied to the totally symmetric Weyl spinor, each $\xi^{(i)}_A$ corresponds to one of the \textit{principal null directions} (PND) familiar from the tensorial version of the Petrov classification. Specifically, the product $\xi^A\bar{\xi}^{A'}\sigma^\mu_{AA'}=k^\mu$ is a principal null vector. The Weyl spinor has at most 4 distinct principal null spinors with corresponding principal null directions, and the Petrov classification merely classifies spacetimes based upon the alignment of these directions. Explicitly,

 \begin{center}
 \begin{tabular}{|c|c|c|}
\hline
 Type & Description & Alt Name \\
 \hline
 I & No PNDs aligned & $\{$1111$\}$ \\
 II & Two PNDs aligned & $\{$211$\}$\\
 D & 2 sets of two PNDs aligned & $\{$22$\}$\\
 III & Three PNDs aligned & $\{$31$\}$\\
 N & Four PNDs aligned & IV or $\{$4$\}$\\
 O & Weyl tensor vanishes & $\{$-$\}$\\
 \hline
\end{tabular}
 \end{center}
Any spacetime that has at least one degenerate PND is called \textit{algebraically special}. For type-D spacetimes specifically, when one chooses the basis spinors to be aligned with the principal null spinors, it becomes evident from the basis expansion of \eqref{basisweyl} that we may always choose:
\begin{equation}
\Psi_0 = \Psi_1 = \Psi_3 = \Psi_4 = 0,\ \ \ \ \Psi_2 \neq 0.
\end{equation}
Notably, this allows us to write the Weyl spinor in the form
\begin{equation}
\Psi_{ABCD} = 6\Psi_2 o_{(A}\iota_B o_{C}\iota_{D)}. \label{weylTypeD}
\end{equation}

\subsection{Maxwell's equations in spinor form}

Crucially to writing Maxwell's equations in spinor form, any antisymmetric rank-2 tensor $F_{\mu\nu}$ can be decomposed in a manner akin the Weyl tensor. Specifically, we can always write,
\begin{equation}
F_{\mu\nu}\sigma^{\mu}_{AA'}\sigma^\nu_{BB'} = F_{AA'BB'} = F_{[AB]A'B'} + F_{AB[A'B']}.
\end{equation}
This is where the simplicity of the spinor formalism shines. For any $D$-dimensional vector space, the maximum rank of any totally antisymmetric tensor is rank-$D$, and all rank-$D$ totally antisymmetric tensors are proportional. Therefore if $D=2$, all antisymmetric rank-2 spinors are proportional to the antisymmetric tensor $\epsilon_{AB}$, and so we must have
\begin{equation}
 F_{\mu\nu}\sigma^{\mu}_{AA'}\sigma^\nu_{BB'} = f_{AB}\epsilon_{A'B'} + \overline{g}_{A'B'}\epsilon_{AB}, \label{MaxwellGen}
\end{equation}
for some $f_{AB}$ and $\overline{g}_{A'B'}$, with $f_{AB} = f_{(AB)}$ and $\overline{g}_{A'B'} =\overline{g}_{(A'B')}$ following from the antisymmetry of $F_{\mu \nu}$, and $\overline{f}_{A'B'} = \overline{g}_{A'B'}$ for any real $F_{\mu \nu}$. Thus, the six real degrees of freedom are contained within the three complex degrees of freedom in $f_{AB}$. By the same argument as \eqref{symmspinor}, any symmetric $f_{AB}$ may be decomposed into a symmetrized product of 1-spinors,
\begin{equation}
 f_{AB} = \alpha_{(A}\beta_{B)}
\end{equation}
where $\alpha^A$ and $\beta^B$ are the principal null spinors of $f_{AB}$. So, one may also classify electromagnetic spinors via the coincidence of their PNDs.

Any rank-2 tensor $F_{\mu \nu}$ can be decomposed into self-dual ($\star F^+_{\mu\nu} = iF^+_{\mu\nu}$) and anti-self-dual ($\star F^-_{\mu\nu} = -iF^-_{\mu\nu}$) pieces under the Hodge-star operation. These pieces are written naturally in the spinor language as follows:
\begin{equation}
 F^-_{\mu\nu}\sigma^{\mu}_{AA'}\sigma^\nu_{BB'} = f_{AB}\epsilon_{A'B'} \qquad F^+_{\mu\nu}\sigma^{\mu}_{AA'}\sigma^\nu_{BB'} =\overline{g}_{A'B'}\epsilon_{AB}. \label{selfdualspin}
\end{equation}
The self-dual and anti-self-dual portions have utility in reducing Maxwell's equations. Consider Maxwell's equations with the inclusion of magnetic charge (meaning nonzero Bianchi identity) represented in 4-vector form as
\begin{equation}
\nabla_\mu F^{\mu\nu} = J_{\text{e}}^\nu,\ \ \ \ \nabla_{\mu} \tilde{F}^{\mu\nu} = J_{\text{m}}^{\nu},\label{maxwell}
\end{equation}
where $\tilde{F}$ is the Hodge dual of $F$, $J_{\text{e}}^\nu$ is the electric 4-current and $J_{\text{m}}^\nu$ is the magnetic 4-current. In order to translate curved-space differential equations into spinor form, we must consider the covariant derivative in spinor form,
\begin{equation}
 \sigma^{\mu}_{AA'}\nabla_\mu = \nabla_{AA'},\label{spincovd}
\end{equation}
whose action on spinor components involves the spin coefficients from the NP formalism $\tensor{\gamma}{_{AA'C}^B}$:
\begin{equation}
 \nabla_{AA'}\kappa^B = \partial_{AA'}\kappa^B + \tensor{\gamma}{_{AA'C}^B}\kappa^C
\end{equation}
By combining the simple form of Maxwell's equation in the language of differential forms,
\begin{equation}
 \mathbf{d}F^- = \frac{1}{2}i \star J,
\end{equation}
with the spinor covariant derivative \eqref{spincovd} and the anti-self-dual form in spinor language \eqref{selfdualspin}, we arrive at one equation containing all of Maxwell's equations in spinor form:
\begin{equation}
 \nabla^A_{B'}f_{AB} = \frac{1}{2}J_{BB'} = \frac{1}{2}(J_{\text{e}}^\mu + iJ_{\text{m}}^\mu)\sigma_{\mu BB'}. \label{MaxwellSpin}
\end{equation}
Thus, the real and imaginary parts of $J_{BB'}$ split into the electric and magnetic 4-currents. Duality rotations, where one rotates electric and magnetic 4-currents into each other via
\begin{equation}
 \begin{pmatrix} F^{\mu \nu}\\\tilde{F}^{\mu \nu} \end{pmatrix} \to \begin{pmatrix}
 \cos(\theta) & -\sin(\theta)\\ \sin(\theta) & \cos(\theta)\end{pmatrix} \begin{pmatrix} F^{\mu \nu}\\\tilde{F}^{\mu \nu} \end{pmatrix} \implies \begin{pmatrix} J_{\text{e}}^\mu\\J_{\text{m}}^\mu \end{pmatrix} \to \begin{pmatrix}
 \cos(\theta) & -\sin(\theta)\\ \sin(\theta) & \cos(\theta)\end{pmatrix} \begin{pmatrix} J_{\text{e}}^\mu\\J_{\text{m}}^\mu \end{pmatrix},
\end{equation}
%(hence leaving the vacuum equations invariant),
also have a very simple interpretation on the spinor side as a \textit{complex phase rotation} of $f_{AB}$:
\begin{equation}
 \nabla^A_{B'}(e^{i\theta}f_{AB}) = \frac{1}{2}e^{i\theta}J_{BB'} = \frac{1}{2}\left[ \big( J_{\text{e}}^\mu \cos(\theta) - J_{\text{m}}^\mu \sin(\theta) \big) + i \big(\cos(\theta) J_{\text{m}}^\mu + J_{\text{e}}^\mu \sin(\theta) \big) \right]\sigma_{\mu BB'} \label{spinDuality}
\end{equation}

\section{Backgrounds in the classical double copy}\label{sec:structureCDC}

Now equipped with the spinorial technology to proceed, we begin by presenting the Kerr--Schild double copy, demonstrating that it may be interpreted on the ``base" metric flat background or the original curved background metric. We then provide an overview of the Weyl double copy prescription, emphasizing that the first such instance of this relation placed ``single-copies" on the full curved background metric, a feature which continues to hold true with sources. We then provide a precise procedure from which one can convert from curved to flat single-copy backgrounds, emphasizing that Kerr--Schild metrics hold special properties in this regard.

\subsection{The Kerr--Schild double copy}\label{KSDC}

Consider a metric of the form:
\begin{equation}
 g_{\mu \nu} = \eta_{\mu \nu} + \psi k_\mu k_\nu
\end{equation}
with a flat Minkowski base metric $\eta_{\mu \nu}$, some scalar function $\psi = \psi(x^\mu)$, and null vector $k_\mu$. Typically a Cartesian base metric is chosen, although this argument also holds when generalizing to other flat base metrics (like Minkowski in cylindrical or spherical coordinates). It can be readily shown \cite{TAUB1981326} that if $k^\mu$ is null, geodesic, or shear-free with respect to $g_{\mu \nu}$, then it must also be null, geodesic, or shear-free with respect to the base metric (regardless of base metric curvature). We also note that the vacuum Einstein equations \textit{imply} that $k^\mu$ must be geodesic and shear-free \cite{Easson:2023dbk}. When matter is present, one can show that \cite{Stephani:2003tm} a Kerr--Schild null vector $k_\mu$ is geodesic if and only if the energy-momentum tensor satisfies
\begin{equation}
 T_{\mu \nu}k^\mu k^\nu = 0.
\end{equation}
Furthermore, the same geodesic null vector $k_\mu$ is a (multiple) principal null direction of the Weyl tensor, meaning all Kerr--Schild spacetimes are algebraically special (not Type I). A remarkable property of geodesic Kerr--Schild spacetimes is that the mixed-index Ricci tensor is linear in the perturbation $\psi$ \cite{TAUB1981326, Stephani:2003tm}, a property that will play great importance in the flat background interpretation of the classical double copy:
\begin{equation}
 \tensor{R}{^\mu_\nu} = \frac{1}{2}\eta^{\mu \alpha}\eta^{\beta \gamma}\partial_{\beta}\left[\partial_{\alpha}(\psi k_\gamma k_\nu) + \partial_{\nu}(\psi k_\gamma k_\alpha) - \partial_{\gamma}(\psi k_\alpha k_\nu) \right]. \label{linearKS}
\end{equation}
Here, $\eta^{\mu \alpha}$ is the inverse Minkowski metric. Note that more general (nonflat) base metrics will include terms proportional to the base metric's Ricci tensor. At this point, it is typically taken that the Kerr--Schild metric is \textit{stationary} in these coordinates, so assuming the timelike Killing vector is in the direction $\partial_0$, we have $\psi$ and $k^\mu$ with no explicit dependence on time. Coupled with the freedom to rescale any null vector such that $k_0 = 1$, this allows us to arrive at the Kerr--Schild double copy \cite{Monteiro:2014cda},
\begin{equation}
 \tensor{R}{^\mu_0} = -\frac{1}{2}\partial_{\beta}\left[\partial^{\beta}(\psi k^\mu) -\partial^{\mu}(\psi k^\beta) \right] \equiv -\frac{1}{2}\partial_{\beta}F_{\text{flat}}^{\beta \mu}, \label{KSflatMixedRicci}
\end{equation}
where $\partial^{\beta} = \eta^{\beta \alpha}\partial_{\alpha}$ is raised with respect to the background metric. We have therefore obtained the single-copy correspondence for Kerr--Schild spacetimes (subject to the above conditions); from the metric ``perturbation" $\psi k_\mu k_\nu$, an electromagnetic potential $A_\mu \equiv \psi k_\mu$ that satisfies Maxwell's equations on the flat ``base metric" background can be constructed. With these conventions, the electromagnetic sources satisfy $J_{\text{e}}^\mu = -2\tensor{R}{^\mu_0}$. When in vacuum, it has been demonstrated \cite{Easson:2023dbk} the Kerr--Schild double copy need not be stationary and can be formulated equivalently to the (still vacuum) Weyl double copy.

However, we note that it is a \textit{choice} to express the Kerr--Schild source $\tensor{R}{^\mu_0}$ in terms of the Minkowski metric $\eta^{\mu \alpha}$. Using the identity \cite{TAUB1981326}:
\begin{equation}
 \partial_\mu k_\nu = \nabla_\mu k_\nu - \frac{1}{2}k_\mu k_\nu k^\sigma \partial_\sigma \psi
\end{equation}
we immediately see that the $F_{\mu \nu}$ constructed from either covariant derivative yields the same electromagnetic field strength,
\begin{equation}
 F^{\text{flat}}_{\mu \nu} = \partial_\mu (\psi k_\nu) - \partial_\nu (\psi k_\mu) = \nabla_\mu (\psi k_\nu) - \nabla_\nu (\psi k_\mu) = F_{\mu \nu}, \label{flatKSMax}
\end{equation}
where we take $F_{\mu \nu}$ to have curved space indices. That this is true implies the Christoffel contractions with $F^{\mu \nu}$ are summed to zero, so no further dependence of the electromagnetic tensor on the spacetime is introduced through the covariant derivative. We can further demonstrate that the curved space $F_{\mu \nu}$ satisfies the same relation \eqref{KSflatMixedRicci} on the original curved background. This follows from the simple property that the volume element is invariant under a Kerr-Schild transformation, implying in this case the determinant is unchanged, simplifying the following general relation for antisymmetric tensors:
\begin{equation}
 \nabla_\mu F^{\mu \nu} = \frac{1}{\sqrt{|g|}}\partial_\mu \left(\sqrt{|g|} F^{\mu \nu} \right) = \partial_\mu F^{\mu \nu}.
\end{equation}
This implies that \eqref{KSflatMixedRicci} can be equivalently expressed as
\begin{equation}
 \tensor{R}{^\mu_0} = -\frac{1}{2}\nabla_{\beta}\left[\nabla^{\beta}(\psi k^\mu) -\nabla^{\mu}(\psi k^\beta) \right] \equiv -\frac{1}{2}\nabla_{\beta}F^{\beta \mu}, \label{KScurved}
\end{equation}
whereby under the same stationary assumptions, we arrive at a Kerr--Schild single copy on the original curved background, with the same electromagnetic potential $A_\mu = \psi k_\mu$, again with the Christoffel contractions summing to zero. In a sense we have performed the derivation backwards; the curvature dependent terms from the covariant derivative drop out in the equations, leading to the equations interpretable on a flat background. Therefore for geodesic stationary Kerr--Schild spacetimes, the single-copy electromagnetic potential identification $A_\mu = \psi k_\mu$ placed on the original curved background will emerge as equivalently living on the flat base metric background. Structurally, one can view Kerr--Schild coordinates as an ``exact" linearization, with the null and geodesic conditions ensuring that higher order terms of the Kerr--Schild perturbation contract to zero.

Nonetheless, there is a prominent difference between the flat background and original background prescriptions. The flat background metric also implies a wave equation, evident from \eqref{KSflatMixedRicci}:
\begin{equation}
 \tensor{R}{^0_0} = -\frac{1}{2}\partial_{\beta}\partial^{\beta}(\psi k^0). \label{KSzero}
\end{equation}
This provides another striking correspondence to the amplitudes double copy, as this wave equation (in vacuum) is the abelianized biadjoint scalar theory's equation of motion. This scalar theory is the ``zeroth copy." However the curved background does not share this feature, since
\begin{equation}
 \partial_{\beta}\partial^{\beta}(\psi) \neq \nabla_{\beta}\nabla^{\beta}(\psi) =  \partial_{\beta}\partial^{\beta}(\psi) - g^{\alpha \beta}\tensor{\Gamma}{_{\alpha\beta}^\mu}\partial_\mu(\psi),
\end{equation}
which reduces for geodesic and null $k_\mu$ into the form:
\begin{equation}
    \nabla_{\beta}\nabla^{\beta}(\psi) =  \partial_{\beta}\partial^{\beta}(\psi) - (k^\mu \partial_\mu \psi)^2 - \psi (k^\mu \partial_\mu \psi)(\partial_\nu k^\nu).\label{LaplacianCurved}
\end{equation}
This discrepancy between the single and zeroth copy is due to the null and geodesic condition on $k^\mu$, and we reiterate that, upon expanding the covariant derivatives in \eqref{KScurved} in terms of Christoffel symbols, one would arrive at an equivalent expression to \eqref{flatKSMax}. However as can be seen the curved-space wave equation \eqref{LaplacianCurved} does not reduce to the flat space wave equation upon expansion. One can equivalently argue that the flat-space zeroth-copy contains no new information, since the flat-space wave equation is implied by \eqref{KSflatMixedRicci}, whereas the curved-space wave equation akin to \eqref{KSzero} has additional nonvanishing terms, evident from \eqref{KScurved}.

\subsection{The Weyl double copy}\label{sec:WeylDC}

While the Kerr--Schild double copy is a profound example of a double copy structure in classical gravity, its coordinate-dependent structure leaves much to be desired in terms of the general understanding for which a ``single copy" may be contained within GR. In comes the Weyl double copy \cite{Luna:2018dpt}, which is based upon the revelation that all vacuum Type-D spacetimes contain single-copy electromagnetic fields on the same curved background \cite{Walker:1970un,Hughston:1972qf}. Consider a vacuum type-D Weyl spinor, where we have chosen our basis to be aligned with the principal null spinors of $\Psi_{ABCD}$ \eqref{weylTypeD}, and (for convenience) define $\Psi \equiv 6\Psi_2$. Then, the following symmetric spinor satisfies the vacuum Maxwell equations on the same curved background [of the form \eqref{MaxwellGen} and \eqref{MaxwellSpin}],
\begin{equation}
 f_{AB} = \Psi^{2/3}o_{(A}\iota_{B)} \implies \nabla^{AA'}f_{AB} = 0.
\end{equation}
This is a consequence of Einstein's equations in vacuum, since all curvature information is contained in the Weyl tensor. In spinor form, the vacuum Einstein equations are
\begin{equation}
 \nabla^{AA'}\Psi_{ABCD} = \nabla^{AA'}\left[\Psi o_{(A}\iota_Bo_C\iota_{D)} \right] = 0.
\end{equation}
As consequence, the Weyl spinor for a vacuum type-D spacetime can always be rewritten into the suggestive form,
\begin{equation}
 \Psi_{ABCD} = \frac{1}{S}f_{(AB}f_{CD)},
\end{equation}
where $S = \Psi^{1/3}$. This correspondence is defined only up to a constant rescaling by a complex number $\kappa e^{i \theta} \in \mathbb{C}$ for $\kappa, \theta \in \mathbb{R}$ via
\begin{equation}
 f_{AB} \to \kappa e^{i \theta} f_{AB}, \ \ \  S \to \kappa^2 e^{2i \theta} S,
\end{equation}
which has an elegant interpretation in terms of the vacuum Maxwell equations. As previously discussed, a duality rotation, which leaves the vacuum Maxwell equations invariant through rotating the electromagnetic field strength and its Hodge dual into one another, simply corresponds to a complex phase rotation of the electromagnetic spinor \eqref{spinDuality}. The rescaling $\kappa$ can be interpreted as the fact that the vacuum Maxwell equations are scale invariant. %Therefore, this ``single-copy" provides direct correspondence between all other possible vacuum Maxwell theories with the same stru.

The Weyl double copy \cite{Luna:2018dpt} utilizes the fact that the most general type-D vacuum metric, the vacuum Plebanski--Demianski metric \cite{Plebanski:1976gy,Griffiths_Pleb}, can be written in (double) Kerr--Schild form when the spacetime is complexified, which in this case still admits a Kerr--Schild double copy. \cite{Luna:2018dpt} demonstrated that the electromagnetic spinor $f_{AB}$ obtained through the above curved space procedure, when mapped to the Kerr--Schild flat base metric, yields precisely the same correspondence that one would obtain through the Kerr--Schild double copy. That we can choose the original curved background or flat base metric in this case might have been guessed from the aforementioned ambiguity of the Kerr--Schild double copy, where one can interpret the electromagnetic field strength as living on either the flat base metric or original curved background. Rather than proving these statements for double Kerr--Schild metrics like Plebanski--Demianski, we provide a more fundamental approach to understanding this ambiguity in Section \ref{flatback}.

We also make a brief note on a comment from \cite{Luna:2018dpt}, where it was stated that the factor $S$ satisfies a wave equation on both curved and flat background: while this is true on the flat background, it does not generally seem to be true on the original curved background.\footnote{We also note this was never claimed to be true in \cite{Walker:1970un,Hughston:1972qf}. Checking the zeroth copy for the simplest cases, such as Schwarzschild, demonstrates it cannot satisfy a source-free wave equation on the originally curved background.} This matches the same observation in Kerr--Schild coordinates, namely that the zeroth copy $\psi$ only satisfies a wave equation on flat background.

\paragraph{The Weyl double copy with sources}

That the Weyl and Kerr--Schild double copies yield the same result in the case of vacuum type-D spacetimes (demonstrated through Plebanski--Demianski) means one might be able to use the Kerr--Schild source prescription to extend the Weyl double copy to include sources. This is precisely what was done in \cite{Easson:2021asd,Easson:2022zoh}, where one finds that, if a given type-D metric is Kerr--Schild and nonvacuum, the Weyl spinor factorizes into a sum of products,
\begin{equation}
 \Psi_{ABCD} = \sum_n \frac{1}{S_{(n)}}f^{(n)}_{(AB}f^{(n)}_{CD)},
\end{equation}
with associated electromagnetic tensors satisfying
\begin{equation}
 \nabla^{(0)}_{\mu}F^{\mu \nu}_{(n)} = J^\nu_{(n)}.
\end{equation}
Here, $\nabla^{(0)}_{\mu}$ is the covariant derivative over the flat base metric, and the division of the Kerr--Schild source $\tensor{R}{^\mu_0}$ into electromagnetic sources $J^\nu_{(n)}$ is based upon the uniqueness of different sources present in the stress-energy tensor. Alternatively, one can phrase this as a separation based upon the degree of polynomial (with respect to the coordinates) within $\Psi_2$. We emphasize that the more fundamental spinorial reason for this decomposition is as of yet unknown, but it has been proven to be true at linearized level \cite{Armstrong-Williams:2024bog}.

\paragraph{A simple example: Reissner--Nordstr\"om}

To provide an explicit example for such a splitting of the Weyl spinor, consider the Reissner--Nordstr\"om metric in the (spherical) Kerr--Schild form,
\begin{equation}
 g_{\mu \nu} = \eta^{(\text{sph})}_{\mu \nu} + \psi(r)k_\mu k_\nu,
\end{equation}
where $\eta^{(\text{sph})}_{\mu \nu}$ is the flat metric in spherical coordinates $(t,r,\theta,\varphi)$, and
\begin{equation}
 \psi(r) = \frac{2M}{r} - \frac{Q^2}{r^2}, \qquad k_\mu = \delta^t_\mu - \delta^r_\mu.
\end{equation}
Reissner--Nord\"strom is a type-D metric, so we can choose a complex null tetrad with only one nonzero Weyl scalar:
\begin{equation}
 \Psi_2 = \frac{Q^2}{r^4} - \frac{M}{r^3}.
\end{equation}
Applying the sourced Weyl double copy \cite{Easson:2021asd}, one finds that, on the same curved background (or flat base metric due to its Kerr--Schild form), the spinor splits into two terms, each corresponding to one of the terms in $\Psi_2$. The corresponding electromagnetic fields are
\begin{equation}
 f^{(1)}_{AB} = \kappa e^{i \beta} \frac{M}{r^2} o_{(A}\iota_{B)}, \qquad f^{(2)}_{AB} = 4\frac{Q^2}{r^3} o_{(A}\iota_{B)},
\end{equation}
where again the factor $\kappa e^{i \beta}$ can be any complex number so long as $f^{(1)}_{AB}$ satisfies the vacuum equations, and the factor 4 in $f^{(2)}_{AB}$ is such to maintain the Kerr--Schild double copy prescription, i.e.
\begin{equation}
 \nabla_\mu F_{(1)}^{\mu \nu} = 0 \qquad\nabla_\mu F_{(2)}^{\mu \nu} = J_{(e)}^\nu = -2 \tensor{R}{^\nu_0} = 2\frac{Q^2}{r^4}\delta^{\nu}_0,\label{maxwellRN1}
\end{equation}
where $F_{(i)}^{\mu \nu}$ is the tensor analogue of $f^{(i)}_{AB}$.

\subsection{Flat background limits in the classical double copy}\label{flatback}

A natural question to ask is under what circumstances there exists an equivalence between Maxwell's equations on the original curved background and an ``appropriate" flat space. Consider the curved space type-D Weyl double copy, which is a coordinate independent statement, while the flat space interpretation is predicated on placing the coordinates in a Kerr--Schild form, meaning the flat background interpretation requires coordinate dependence to draw the correspondence, the first indication the curved background being more fundamental. Furthermore, the ``appropriate" flat background to place the single copy is determined by the Kerr--Schild form, meaning if one wishes to generalize the classical double copy beyond Kerr--Schild (as in this work), we must lay out an explicit procedure for taking the flat space limit of curved space Maxwell fields.

In this section we adapt an argument from Penrose and Rindler \cite{Penrose_Rindler_1984} demonstrating that the type-D Weyl double copy in its spinorial form generally is constructed on the curved original background, and in specific cases may be smoothly deformed into flat space. We begin by focusing on Kerr--Schild coordinates, with the generalization naturally following.

Consider a smooth family of spacetimes $(\mathcal{M}_\alpha,g(\alpha))$ with constant parameter $\alpha$.
%, such that as $\alpha \to 0$, the spacetime with associated metric continues to satisfy Einstein's equations.
We also stipulate that, as $\alpha \to 0$, the spacetime approaches Minkowski space, meaning $g_{\mu \nu}(\alpha) \to \eta_{\mu \nu}$ as $\alpha \to 0$.
%The parameter $\alpha$ will show up in curvature quantities and the energy-momentum tensor, assuming they do not vanish.
Since our space approaches flat space, curvature quantities tend to zero as $\alpha \to 0$, and in particular the Christoffel symbols vanish in this limit. Notice that a Kerr--Schild perturbation of a Minkowski base metric naturally satisfies this requirement, so long as the perturbation vanishes as $\alpha \to 0$:
\begin{equation}
g_{\mu \nu} = \eta_{\mu \nu} + \psi(\alpha,x^\mu) k_\mu k_\nu \xrightarrow{\alpha \to 0} \eta_{\mu\nu}\ \ \text{if}\ \ \lim_{\alpha\to 0} \psi(\alpha,x^\mu) = 0.
\end{equation}
Now, suppose the functional dependence of the perturbation on $\alpha$ is completely separable from the coordinate dependence:
\begin{equation}
\psi(\alpha,x^\mu) \equiv \mathbf{f}(\alpha)\phi(x^\mu).
\end{equation}
So, $\mathbf{f}(\alpha) \to 0$ as $\alpha \to 0$. Kerr--Schild coordinates have some remarkable properties, as discussed in Section \ref{KSDC}, and of note is the linearity of single-copy quantities in the perturbation. That $\psi(\alpha,x^\mu)$ is separable immediately implies that both the electromagnetic field strength and its source are also separable, with the same functional dependence on $\alpha$:
\begin{align}
F_{\mu \nu} &= \nabla_\mu (\psi k_\nu) - \nabla_\nu (\psi k_\mu) = \mathbf{f}(\alpha)\left[ \nabla_\mu (\phi k_\nu) - \nabla_\nu (\phi k_\mu) \right] \equiv \mathbf{f}(\alpha)\mathcal{F}_{\mu \nu}(x^\mu), \label{adepF}\\
\tensor{R}{^\mu_0} &= -\frac{1}{2}\nabla_{\beta}\left[\nabla^{\beta}(\psi k^\mu) -\nabla^{\mu}(\psi k^\beta)  \right] = -\frac{1}{2}\mathbf{f}(\alpha)\nabla_{\beta}\left[\nabla^{\beta}(\phi k^\mu) -\nabla^{\mu}(\phi k^\beta)  \right] \equiv \mathbf{f}(\alpha)\mathcal{R}^\mu(x^\mu). \label{adepR}
\end{align}
It was demonstrated previously that Kerr--Schild coordinates have equivalent equations for \eqref{adepF} and \eqref{adepR} taking $\nabla_\mu \to \partial_\mu$, which implies the Christoffel contractions from the covariant derivative sum to zero. On the level of Maxwell's equations in curved space, one can demonstrate this fact either explicitly as done before, or by considering the following limiting procedure:
\begin{align*}
\nabla_\mu F^{\mu \nu} = -2 \tensor{R}{^\mu_0}
&\implies \mathbf{f}(\alpha)\nabla_\mu \mathcal{F}^{\mu \nu} = -2 \mathbf{f}(\alpha) \mathcal{R}^\mu\\
&\implies  \nabla_\mu \mathcal{F}^{\mu \nu} = -2 \mathcal{R}^\mu\\
&\xrightarrow{\alpha \to 0} \partial_\mu \mathcal{F}^{\mu \nu} = -2 \mathcal{R}^\mu. \numberthis \label{flatspaceprocedure}
\end{align*}
The last step follows from $\Gamma_{\mu \nu}^\rho(\alpha) \to 0$ as $\alpha \to 0$. Our main takeaway is that, for metrics of a geodesic Kerr--Schild form, the covariant derivative of an electromagnetic field does not introduce any new functional dependence on $\alpha$ into the source $J^\mu$. However, this remarkable fact is not generically true for the curved-background Maxwell equations. 

Generalizing this procedure to include multiple parameters is straightforward. Say that a family of spacetimes is parameterized by multiple parameters $\alpha_n$, where each parameter can be smoothly deformed to zero such that when all $\alpha_n \to 0$, the spacetime becomes Minkowski. In Kerr--Schild form, we take that the perturbation is separable into a sum of multiple parameters $\alpha_n$:
\begin{equation}
    \psi(\alpha_1,\dots,\alpha_n,x^\mu) = \sum_n \mathbf{f}_n(\alpha_n)\phi_n(x^\mu).
\end{equation}
Once again, $F_{\mu \nu}$ and $\tensor{R}{^\mu_\nu}$ being linear in the perturbation $\psi$ ensures the same dependence on parameters $\alpha_n$, and we can extract the dependence on any specific parameter by sending all of the other parameters to zero:
\begin{align}
F_{\mu \nu} &= \sum_n \mathbf{f}_n(\alpha_n) \mathcal{F}^{(n)}_{\mu \nu} \ \ \implies \ \ \lim_{\{\alpha_i \neq \alpha_N\} \to 0} F_{\mu \nu} = \mathbf{f}_N (\alpha_N) \mathcal{F}^{(N)}_{\mu \nu},\\
\tensor{R}{^\mu_0} &= \sum_n \mathbf{f}_n(\alpha_n)\mathcal{R}_{(n)}^{\mu} \ \ \implies \ \ \lim_{\{\alpha_i \neq \alpha_N\} \to 0} \tensor{R}{^\mu_0} = \mathbf{f}_N (\alpha_N) \mathcal{R}_{(N)}^{\mu}. 
\end{align}
Since Maxwell's equations are linear in $F^{\mu \nu}$ and the source $J^\mu$, taking the above limits for each separate parameter $\alpha_N$ and then repeating the sequence from \eqref{flatspaceprocedure} yields equivalent equations on a flat background. These terms can be  resummed to produce the full electromagnetic tensor and source satisfying the flat-space Maxwell equations.

So the question is as follows; generically, for spacetimes without a Kerr--Schild coordinatization, will they satisfy equivalent Maxwell equations on curved and an ``appropriate" flat space? In the case of a single parameter $\alpha$, the answer is straightforward. If the covariant derivative introduces dependence on the parameter $\alpha$, then the limit $\alpha \to 0$ will generically alter the structure of the source $J^\mu \sim \mathcal{R}^\mu$.

We now seek an analogue of this argument for the Weyl double copy. Consider the spinorial version of Maxwell's equations from \eqref{MaxwellSpin},
\begin{equation}
    \nabla^A_{B'}f_{AB} = \frac{1}{2}J_{BB'},
\end{equation}
where $J_{BB'}$ is the spinorial analogue of the complex current $J_\mu = J^{(e)}_\mu + iJ^{(m)}_\mu$. Since $f_{AB}$ in the classical double copy is derived from curvature quantities (namely the Weyl spinor), it will have dependence on $\alpha$. In the case of the type-D Weyl double copy, the Maxwell spinor will necessarily have the form (upon choosing the principal null spinors of the Weyl spinor to be the basis)
\begin{equation}
    f_{AB} = \phi(\alpha,x^\mu) o_{(A}\iota_{B)}. \label{fABwphi}
\end{equation}
Take the functional dependence on $\alpha$ of $\phi(\alpha,x^\mu)$ to be $\mathbf{f}(\alpha)$, such that $\phi(\alpha, x^\mu) = \mathbf{f}(\alpha)\xi(x^\mu)$. We can therefore consider the limit:
\begin{equation}
    \lim_{\alpha \to 0}\nabla^A_{B'}\frac{1}{\mathbf{f}(\alpha)}f_{AB} =\lim_{\alpha \to 0} \nabla^A_{B'}\xi(x^\mu) o_{(A}\iota_{B)} = \lim_{\alpha \to 0} \frac{1}{\mathbf{f}(\alpha)}J_{BB'}. \label{flatSpinLim}
\end{equation}
Note that now, when the limit $\alpha \to 0$ is taken, the covariant derivative will become one on flat space, by assumption. Upon translating back into tensorial form, we see that
\begin{equation}
    \partial_\mu \mathcal{F}^{\mu \nu} = \lim_{\alpha \to 0} \frac{1}{\mathbf{f}(\alpha)}J_{(e)}^\nu,
\end{equation}
with the RHS being dependent on the spacetime, and not necessarily even well defined.

\paragraph{The flat space procedure explicitly (Reissner--Nordstr\"om)}

To again provide an explicit example for clarity, we revisit the Reissner--Nordstr\"om metric previously analyzed through the sourced Weyl double copy in Section \ref{sec:WeylDC}. Consider now the flat limit of the Reissner--Nordstr\"om metric, which can be achieved through letting $M \to 0$ and $Q \to 0$ simultaneously. Recall the electromagnetic spinors:
\begin{equation}
 f^{(1)}_{AB} = \kappa e^{i \beta} \frac{M}{r^2} o_{(A}\iota_{B)}, \qquad f^{(2)}_{AB} = 4\frac{Q^2}{r^3} o_{(A}\iota_{B)}.
\end{equation}
Using our aforementioned prescription for yielding flat space fields (and defining the covariant derivative $\nabla^{(0)}_\mu$ as living on the flat background in spherical coordinates):

\begin{equation}
 \lim_{\{M,Q\} \to 0}\nabla^A_{B'}\frac{1}{M} f^{(1)}_{AB} = 0 \longrightarrow \nabla^{(0)}_\mu \mathcal{F}_{(1)}^{\mu \nu} = 0,
\end{equation}
\begin{equation}
 \lim_{\{M,Q\} \to 0}\nabla^A_{B'}\frac{1}{Q^2} f^{(2)}_{AB} = \lim_{\{M,Q\} \to 0}\frac{1}{Q^2} J_{BB'} \longrightarrow \nabla^{(0)}_\mu \mathcal{F}_{(2)}^{\mu \nu} = 
 \lim_{\{M,Q\} \to 0} \frac{1}{Q^2}J_{(e)}^\nu = 2\frac{1}{r^4}\delta^{\mu}_0,
\end{equation}
where we have made the following definitions:
\begin{equation}
 \mathcal{F}^{\mu\nu}_{(1)} = \frac{1}{M}F_{(1)}^{\mu \nu} \qquad \mathcal{F}^{\mu\nu}_{(2)} = \frac{1}{Q^2}F_{(2)}^{\mu \nu}.
\end{equation}
The single-copy fields found through this flat-space limit satisfy the same Maxwell equations as those found by invoking the Kerr--Schild representation of the Reissner--N\"ordstrom black hole \eqref{maxwellRN1}. In other words, there is consistency between the flat-space limit and the direct construction of single copies on flat space.

\section{The G\"odel universe}\label{sec:godel}

The G\"odel metric describes a rotating dust solution to Einstein's equations \cite{Godel:1949ga}, and an excellent exposition of this solution is given by Hawking and Ellis \cite{Hawking:1973uf}. To construct it, we first consider the Einstein equation (setting $c = 8\pi G_{\text{N}} = 1$) with negative cosmological constant $\Lambda < 0$ and sourced by a dust of density $\rho$: 
\begin{equation}
G^{\mu\nu} + \Lambda g^{\mu\nu} = \rho U^\mu U^\nu.
\end{equation}
Now, we suppose that the density of the dust is of the same order as the cosmological constant (defining an inverse-length scale $\omega \in \mathbb{R}$):
\begin{equation}
\Lambda = -\frac{\omega^2}{2},\ \ \ \ \rho = \omega^2,
\end{equation}
and work in the rest frame of the dust (denoting the time coordinate as $t$):
\begin{equation}
U^{\mu} = \delta^\mu_t.\label{fourvelDust}
\end{equation}
The G\"odel metric, which solves the resulting Einstein equations, is
\begin{equation}
ds_{\text{G}}^2 = -(dt + e^{\omega x} dy)^2 + dx^2 + \frac{1}{2}e^{2\omega x} dy^2 + dz^2.\label{godelmetric}
\end{equation}
The coordinates $(t,x,y,z)$ are real and have dimensions of length. In fact, the G\"odel metric is stationary and homogeneous (thus geodesically complete) and does not even have any singularities. However, it is notoriously plagued by closed timelike curves that reach asymptotic infinity.

We can explicitly construct simple families of such curves in a cylindrical coordinate system. Following \cite{Hawking:1973uf}, we first perform the following transformation of $(t,x,y) \to (\tau,r,\phi)$:
\begin{align}
e^{\omega x} &= \cosh(\omega r) + \sinh(\omega r)\cos\phi,\label{coordcyl1}\\
\omega y e^{\omega x} &= \sqrt{2}\sinh(\omega r)\sin\phi,\\
e^{-\omega r}\tan\frac{\phi}{2} &= \tan\left[\frac{1}{2}\left(\phi + \frac{\omega}{\sqrt{2}}(t-\tau)\right)\right].\label{coordcyl3}
\end{align}
Note that we are still in the rest frame of the dust---the four-velocity $U^\mu \partial_\mu$ transforms from $\partial_t$ to $\partial_\tau$. In addition, the transformed metric is
\begin{equation}
ds_{\text{G}}^2 = -d\tau^2 + dr^2 + \frac{4}{\omega^2} \sinh^2\left(\frac{\omega r}{2}\right) \left[1 - \sinh^2\left(\frac{\omega r}{2}\right)\right] d\phi^2 - \frac{4\sqrt{2}}{\omega}\sinh^2\left(\frac{\omega r}{2}\right) d\tau\,d\phi + dz^2.\label{godelMetricCyl}
\end{equation}
The coordinates satisfy $\tau,z \in \mathbb{R}$, $r \geq 0$, and $\phi \sim \phi + 2\pi$. As shown visually by \cite{Hawking:1973uf} (reproduced in Figure \ref{figs:HawkingEllisGodel}), null geodesics wind around the $r = 0$ axis. Interestingly, this causal structure also includes a null circle (i.e. a closed null curve) on each fixed-$\tau$ slice:
\begin{figure}
\centering
\begin{tikzpicture}

\draw[->] (-2,1) to (-2,2);
\node at (-2.2,2) {$\tau$};

\draw[->] (0,2.25) to (1,2.25);
\node at (1,2.025) {$r$};

\draw[-,blue,dashed,thick] (0,0) circle (2*1.5 and 0.25*1.5);
\draw[-,blue,dashed,thick] (0,0) circle (2*2 and 0.25*2);

\draw[-,red] (0,-2) .. controls (0.5,-0.8) and (1.5,0) .. (2,0);
\draw[-,red] (0,-2) .. controls (-0.5,-0.8) and (-1.5,0) .. (-2,0);

\draw[-,fill=red!10,draw=none] (0,0) circle (2 and 0.25);

\draw[-,dashed] (0,-2.5) to (0,2.5);

\draw[-,red,dashed,thick] (0,0) circle (2 and 0.25);

\node[red] at (0,2) {$\bullet$};
\node[red] at (0,-2) {$\bullet$};

\draw[-,dashed,red] (0,-2) .. controls (0.5,-0.8) and (1.5,0) .. (2,0);
\draw[-,dashed,red] (0,-2) .. controls (-0.5,-0.8) and (-1.5,0) .. (-2,0);

\draw[-,red] (0,2) .. controls (0.5,0.8) and (1.5,0)  .. (2,0);
\draw[-,red] (0,2) .. controls (-0.5,0.8) and (-1.5,0) .. (-2,0);

\draw[-,red] (-1.375,0.25) arc (220:340:1.5 and 0.5);
\draw[-,red] (-1.05,0.5)  arc (230:350:1.15 and 0.4);
\draw[-,red] (-0.75,0.8) arc (250:360:0.95 and 0.3);
\draw[-,red] (-0.475,1.15) arc (270:360:0.75 and 0.3);
\draw[-,red] (-0.22,1.55) arc (300:360:0.44 and 0.44);

\draw[-,red,thick] (0,-0.25) arc (270:141:2 and 0.25);
\draw[-,red,thick] (0,-0.25) arc (270:399:2 and 0.25);

\draw[-,red] (0,-2) arc (200:100:0.25 and 0.3);
\draw[-,red] (-0.22,-1.55) .. controls (-0.11,-1.4) and (0.27,-1.3) .. (0.475,-1.15);
\draw[-,red] (-0.475,-1.15) .. controls (-0.2375,-1) and (0.375,-0.95) .. (0.75,-0.8);
\draw[-,red] (-0.75,-0.8) .. controls (-0.375,-0.5) and (0.6,-0.6) .. (1.05,-0.5);
\draw[-,red] (-1.05,-0.5) arc (110:90:3.1 and 4.1) arc (270:290:4.3 and 7.1);

\draw[-,blue,thick] (0,-0.25*1.5) arc (270:114:2*1.5 and 0.25*1.5);
\draw[-,blue,thick] (0,-0.25*1.5) arc (270:425.5:2*1.5 and 0.25*1.5);

\draw[-,blue,thick] (0,-0.25*2) arc (270:105:2*2 and 0.25*2);
\draw[-,blue,thick] (0,-0.25*2) arc (270:435:2*2 and 0.25*2);

\draw[-,red] (0,2) .. controls (0.5,0.8) and (1.5,0)  .. (2,0);
\draw[-,red] (0,2) .. controls (-0.5,0.8) and (-1.5,0) .. (-2,0);
\draw[-,red] (-1.05,0.5)  arc (230:350:1.15 and 0.4);

\node at (0,2.75) {$r = 0$};

\draw[->] (4.5,0) arc (0:30:4.5 and 4.5/8);
\draw[-] (4.5,0) arc (0:-30:4.5 and 4.5/8);

\node at (3.9,0.525) {$\phi$};

\draw[->,white] (-4.5,0) arc (180:150:4.5 and 4.5/8);
\draw[-,white] (-4.5,0) arc (180:210:4.5 and 4.5/8);

\node[white] at (-3.9,0.525) {$\phi$};

\end{tikzpicture}
\caption{A visual depiction of the causal structure of the G\"odel metric, also seen in Figure 31 of \cite{Hawking:1973uf} and with the $z$ direction suppressed. The red curves wrapping around the central $r = 0$ axis are null geodesics. These geodesics constitute a diamond whose central fixed-$\tau$ slice (the red circle) is a closed null curve $r = r_{\text{CNC}}$ \eqref{cncEq}. The larger blue circles $r > r_{\text{CNS}}$ are closed timelike curves on the $\tau = 0$ slice.}
\label{figs:HawkingEllisGodel}
\end{figure}
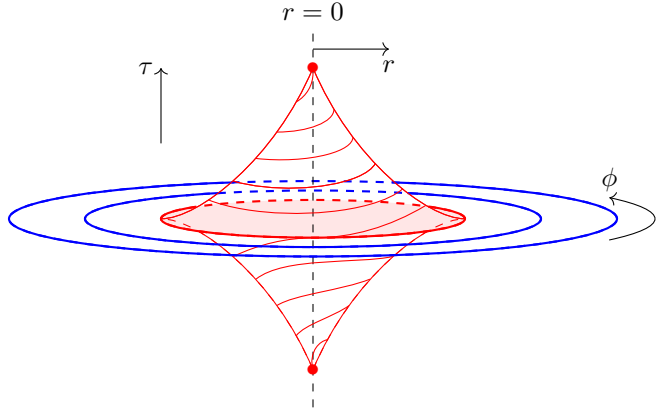
\begin{equation}
r = r_{\text{CNC}} \equiv \frac{2}{\omega}\log(1 + \sqrt{2}).\label{cncEq}
\end{equation}
which is merely the $r$-coordinate where the sign of $d\phi^2$ flips to become a timelike coordinate. The identification $\phi \sim \phi + 2\pi$ therefore ensures for $r > r_{\text{CNC}}$ there exists a family of ``circular" closed timelike curves. Not only are these closed timelike curves on every fixed-$\tau$, fixed-$z$ slice, but they also reach asymptotic infinity without being obstructed by a horizon! The specific value of this radius is more easily parameterized via the metric transformation outlined in \cite{Kajari2004}, whereby the radial coordinate can be recast as:
\begin{equation}
    \rho \equiv \frac{2}{\omega}\sinh\left(\frac{\omega r}{2}\right)
\end{equation}
making the new metric:
\begin{equation}
ds_{\text{G}}^2 = -d\tau^2 + \left[1 + \left(\frac{\omega \rho}{2} \right)^2 \right]^{-1}d\rho^2 +  \rho^2 \left[1 - \left(\frac{\omega \rho}{2}\right)^2\right] d\phi^2 - \sqrt{2}\omega \rho^2 d\tau\,d\phi + dz^2.\label{godelMetricCylNew}
\end{equation}
with closed timelike curves (of constant $\tau, r, z$) now evident for:
\begin{equation}
    \rho > \rho_{\text{CNC}} \equiv \frac{2}{\omega}\label{rhocnc}
\end{equation}
Note that the G\"odel metric actually has eternal closed timelike curves---that is, they cross every event in the spacetime. One way to see this is to exploit the stationary and homogeneous nature of the planar metric \eqref{godelmetric}, which together allow us to define the rotational axis 
of \eqref{godelMetricCyl} along any flow line of the matter \cite{Hawking:1973uf}. However to be clear, we will be working within a single cylindrical coordinate frame when formulating our putative single-copy fields.

\paragraph{The characteristic properties of G\"odel}

Since ultimately we are \textit{constructing} single copy solutions rather than following a well-defined procedure, we identify the characteristic properties of the G\"odel universe that we seek a correspondence for:
\begin{enumerate}
 \item[(I)] The spacetime is stationary and axisymmetric;
 \item[(II)] The spacetime is homogeneous;
 \item[(III)] About any point there exists a non-vanishing constant vorticity vector, representing the dust source rotating about itself in any selected origin (``everywhere rotating"); and,
 \item[(IV)] In the coordinates manifesting axisymmetry, circular closed timelike curves exist outside a defined radius. Homogeneity implies that this should be a feature around any selected origin.
\end{enumerate}
Note that the symmetries enumerated in (I) being symmetries of the single copy follows from both the Kerr--Schild and the Weyl constructions. If the perturbation $\psi k_\mu k_\nu$ or the Weyl scalar $\Psi_2$ are independent of a specific coordinate, so too will the single copies constructed from them.

\subsection{The Weyl spinor of G\"odel}

Now we bring in the technology of the spinor formalism. In doing so, it is easier to start with the coordinate frame of \eqref{godelMetricCylNew}. First, we note that the G\"odel metric \eqref{godelMetricCylNew} is known to be of type D in the Petrov classification. We can see this from the following complex null tetrad satisfying \eqref{cplxnullgeneric}:
\begin{align}
l^\mu &= \frac{1}{\sqrt{2}}(\delta^\mu_\tau + \delta^\mu_z),\label{cplxg1}\\
n^\mu &= \frac{1}{\sqrt{2}}(\delta^\mu_\tau - \delta^\mu_z),\label{cplxg2}\\
m^\mu &= \frac{\sqrt{g_{\rho \rho}}}{2}\left(-i \omega \rho \delta^\mu_\tau + \frac{1}{2 \sqrt{2} g_{\rho \rho}}\delta^\mu_\rho + \frac{i \sqrt{2}}{\rho}\delta^\mu_\phi \right)\label{cplxg3}
\end{align}
For this choice, we get that the Weyl scalars are
\begin{equation}
\Psi_0 = \Psi_1 = \Psi_3 = \Psi_4 = 0,\ \ \ \ \Psi_2 = -\frac{\omega^2}{6},
\end{equation}
which falls into the type-D class. Furthermore with this tetrad, the resulting Weyl spinor is indeed in the form needed for the Weyl double copy procedure \cite{Luna:2018dpt},
\begin{equation}
\Psi_{ABCD} = -\omega^2 o_{(A}\iota_{B}o_{C}\iota_{D)}.\label{weylSpinorGodel}
\end{equation}

\subsection*{Flat limit of G\"odel}\label{Godelflat}

%\sanjit{need to make sure that the vierbein we write here is compatible with the above null tetrad when switched back to an orthonormal spacelike basis}

Before we use the Weyl spinor, it helps to emphasize that a flat limit background metric can be obtained through a limiting procedure as discussed in Section \ref{flatback}. Specifically, consider taking $\omega \to 0$ in \eqref{godelMetricCylNew}, which corresponds to sending both the cosmological constant and the dust density to $0$ simultaneously. The metric becomes
\begin{equation}
\lim_{\omega \to 0} ds_{\text{G}}^2 = -d\tau^2 + d\rho^2 + \rho^2 d\phi^2 + dz^2.\label{cylBackground}
\end{equation}
This is just Minkowski spacetime in the usual cylindrical coordinates. While it may not be immediately obvious that the flat limit of the Cartesian-like coordinates \eqref{godelmetric} is Minkowski space, the flat limit becomes:
\begin{equation}
\lim_{\omega \to 0} ds_{\text{G}}^2 = -(dt + dy)^2 + dx^2 + \frac{1}{2}dy^2 + dz^2.\label{backgroundPlanar}
\end{equation}
which is Minkowski spacetime in atypical coordinates. By applying the coordinate transformations $t \to t' - \sqrt{2}y'$ and $y \to \sqrt{2}y'$, we get the usual Cartesian metric $-dt'^2 + dx^2 + dy'^2 + dz^2$. Finally, note that by using the identity $\sinh(\alpha x)/\alpha \underset{\alpha \to 0}{\longrightarrow} x$, we yield the same flat limit \eqref{cylBackground} for the alternative cylindrical coordinates \eqref{godelMetricCyl}.

\subsection{G\"odel is not (geodesic) Kerr--Schild}\label{notKS}

Before jumping into the novel approaches for bootstrapping a single copy, we first must confirm that the previously well-defined approaches for type-D classical double copies with sources are not applicable, specifically that it is not of (geodesic) Kerr--Schild form. % or electro-vacuum. The latter is immediately apparent from the spacetime having a nonvanishing Ricci scalar (since the Maxwell stress-energy tensor is traceless). 
Recall that the geodesic condition on the Kerr--Schild perturbation was required for the form of the mixed Ricci tensor \eqref{linearKS}, which in turn is the basis for the Kerr--Schild double copy.

In order to demonstrate that the metric cannot be placed into (geodesic) Kerr--Schild form, we reiterate the relevant conditions on the stress-energy tensor from Section \ref{KSDC}; a Kerr--Schild null vector $k_\mu$ is geodesic if and only if the energy-momentum tensor satisfies
\begin{equation}
 T_{\mu \nu}k^\mu k^\nu = 0. \label{KScondition}
\end{equation}
Furthermore, the same geodesic null vector $k_\mu$ is a (multiple) principal null direction of the Weyl tensor. Recall that, for type-D metrics, we can choose our basis spinors to lie along the principal null spinors. This yields the following Weyl spinor:
\begin{equation}
 \Psi_{ABCD} = 6\Psi_2 o_{(A}\iota_Bo_C \iota_{D)}.
\end{equation}
So, our basis spinors from \eqref{weylSpinorGodel} are principal null spinors.

The Kerr--Schild double copy requires the null vector $k_\mu$ to be geodesic. And so by the above theorems, if G\"odel can be placed into Kerr--Schild form, then the null vector of the perturbation must also be a principal null direction. There are only two principal null directions for G\"odel, proportional to either
\begin{equation}
o^Ao^{A'}\sigma^\mu_{AA'} = \ell^\mu,
\end{equation}
or
\begin{equation}
\iota^A\iota^{A'}\sigma^\mu_{AA'} = n^\mu.
\end{equation}
Therefore for G\"odel to have a chance to be Kerr--Schild, \eqref{KScondition} must be satisfied for one of the above null vectors. One can readily check that, to the contrary,
\begin{equation}
T_{\mu \nu}\ell^\mu \ell^\nu = \frac{\omega^2}{2} = T_{\mu \nu}n^\mu n^\nu.
\end{equation}
Thus, the G\"odel metric cannot be written in geodesic Kerr--Schild form.

\section{Constructing single-copy solutions}\label{sec:SingleCopy}

Since we have demonstrated that the Kerr--Schild double copy is not applicable to our spacetime, we instead choose the Weyl double copy as the foundation from which we will construct possible single-copies. With the Weyl spinor identification \eqref{weylSpinorGodel}, we investigate reasonable single-copies one could construct without an a priori well-defined procedure for identifying sources on the single-copy side. We should note there have been previous attempts to treat ``G\"odel-type" solutions as double copies \cite{G_rses_2005,G_rses_2005_II, G_rses_2018}. This work provides a fascinating correspondence between a Maxwell single-copy theory and Einstein-Maxwell-dilaton-3-form field theory, but the explicit constructions are unfortunately restricted to either $D \geq 6$ or metrics with constant determinant, neither of which is applicable to the 4d G\"odel metric in its original form.

Motivated by the fact that a flat-space Maxwell theory can always be achieved (Section \ref{flatback}) but not vice versa, we begin our analysis of possible single copies by constructing them on the original curved space, noting we can always take a limiting procedure $\omega \to 0$ to obtain a flat background. We find that generically this limiting procedure will result in vanishing sources on the flat background. We compare our single-copy to the defining properties of the spacetimes outlined in Section \ref{sec:godel}. We then demonstrate that, if one truly wants a sourced single copy on flat space, leveraging symmetries is the only reasonable step that can be taken. This yields a rather unsatisfying theory, missing the defining properties of the spacetime.

\subsection{The curved space double copy on G\"odel}

Recall that for the sourced Weyl double copy outlined in Section \ref{sec:WeylDC}, the Weyl spinor generically decomposes into a sum of products, in accordance with some decomposition of $\Psi_2$ \cite{Easson:2021asd,Easson:2022zoh}. However since our Weyl scalar $\Psi_2 \sim \omega^2$ is a constant, 
%unless new functional dependence is introduced though some (unknown as of yet) procedure,
a decomposition into a sum of terms is not evident.

We begin by assuming that no new coordinate dependence is introduced into the single copy, later reevaluating this assumption. In this case, a viable Weyl double copy is achieved by multiplying the electromagnetic spinor of the form \eqref{fABwphi} by a complex constant $\kappa e^{i\theta}$:
\begin{equation}
 f_{AB} = i \omega \kappa e^{i \theta} o_{(A}\iota_{B)}, \qquad S = (\kappa e^{i \theta})^2. \label{genfABnofunc}
\end{equation}
However, it is immediate that we are foregoing a sourced zeroth copy, since $\square S = 0$. That said, since the curved space zeroth copy is generically unknown, we do not consider this a detriment to this correspondence, and we will find that our proposed single copy on flat space is consistent with the flat space zeroth copy. Translating this choice into tensor form yields the electromagnetic fields,
\begin{equation}
 F_{\mu \nu} = \omega \kappa \begin{pmatrix}
0 & 0 & 0 & - \sin(\theta)\\
0 & 0 & \rho \cos(\theta) & 0\\
0 & - \rho \cos(\theta) & 0 & -\frac{1}{\sqrt{2}}\omega \rho^2 \sin(\theta)\\
 \sin(\theta) & 0 & \frac{1}{\sqrt{2}}\omega \rho^2 \sin(\theta) & 0
\end{pmatrix}, \label{Fdd}
\end{equation}
whose Hodge dual is
\begin{equation}
 \tilde{F}_{\alpha\beta} = \omega \kappa\begin{pmatrix}
0 & 0 & 0 & \cos(\theta)\\
0 & 0 & \rho \sin(\theta) & 0\\
0 & - \rho \sin(\theta) & 0 & \frac{1}{\sqrt{2}}\omega \rho^2 \cos(\theta)\\
-\cos(\theta) & 0 & -\frac{1}{\sqrt{2}}\omega \rho^2 \cos(\theta) & 0
\end{pmatrix}.\label{HDFdd}
\end{equation}
Since the G\"odel metric is both nonvacuum and non-Kerr--Schild, we do not know a priori what the sources should be on the single-copy side. Nonetheless, the above field tensors are consistent with the following currents:
\begin{equation}
 J_{\text{(e)}}^\alpha = \begin{pmatrix}
-\sqrt{2}\kappa \omega^2 \cos(\theta)\\0\\0\\0
\end{pmatrix},\ \ J_{\text{(m)}}^{\alpha} = \begin{pmatrix}
-\sqrt{2}\kappa \omega^2 \sin(\theta)\\0\\0\\0
\end{pmatrix}. \label{genKScurrents}
\end{equation}
The electric and magnetic fields are of course observer-dependent. A natural observer could be the one co-moving with the dust which sources the spacetime, which is equivalent to choosing $u^\mu = \delta^\mu_\tau$. The resulting fields are
\begin{equation}
 E^\mu = \tensor{F}{^\mu_\nu}u^\nu = \kappa \omega \sin(\theta)\delta^\mu_z, \quad B^\mu = \tensor{\tilde{F}}{^\mu_\nu}u^\nu = -\kappa \omega \cos(\theta)\delta^\mu_z,
\end{equation}
which are constant electric and magnetic fields. The sources above may seem very confusing, as a purely electric charge source implies a constant magnetic field. In a flat spacetime, this is nonsensical, but in a curved spacetime such as G\"odel the electromagnetic sources are formed by applying a covariant derivative, and so the curvature of spacetime itself may warp the usual identification of how sources relate to fields. In G\"odel, there is a conceptual way to understand this correspondence: due to the ``everywhere-rotating" nature of G\"odel, any two worldlines of matter (and hence of electromagnetic charge) twist around one another. So a constant charge density at each point will imply constant rotation of electric charge about the origin, hence the emergence of a purely magnetic field for the observer at the origin. Note as well the difference in powers of $\omega$ between $F_{\mu \nu}$ and $J^\mu$; an additional factor of $\omega$ enters from the covariant derivative.

From these formulas we also reiterate a well-known fact about complex rotations to $f_{AB}$; multiplication by a complex phase $f_{AB} \to e^{i \theta} f_{AB}$ is equivalent to performing a \textit{duality rotation}, in effect swapping electric and magnetic charges. In particular if we want zero magnetic charge, a natural choice of complex phase is apparent: $\theta = \{0,\pi\}$, leaving only $\kappa$ undetermined. 

\subsection*{Kerr--Schild sources}

To reiterate, we have seen that enforcing no magnetic charges does not completely determine the electromagnetic source, as there is a leftover scaling ambiguity $\kappa$. It is here we turn to other work on sourced classical double copies to write an ansatz. Despite the G\"odel universe not having a Kerr--Schild expansion, one might ask if the source prescription from the Kerr--Schild double copy (for some normalized timelike Killing vector $u^\mu$),
\begin{equation}
 J^\mu = -2 \tensor{R}{^\mu_\nu}u^\nu, \label{KSsources}
\end{equation}
is a more general property of double copies and hence still applicable. The G\"odel metric does have a timelike Killing vector $u^\mu = \partial_t$, and so one finds:
\begin{equation}
 \tensor{R}{^\mu_\nu}u^\nu = \begin{pmatrix}
 - \omega^2\\0\\0\\0
 \end{pmatrix} 
\end{equation}
We can see that the identification \eqref{KSsources} is valid when the choices $\kappa = \sqrt{2}$ and $\theta = \pi$ are made. These choices imply that
\begin{equation}
 f_{AB} = -i \omega \sqrt{2} o_{(A}\iota_{B)} \qquad S = 2. \label{KSfAB}
\end{equation}
It is nontrivial that the identification \eqref{KSsources} can be made at all, as we will see when investigating sources on a flat background in Section \ref{flatspacesec}.

\subsection*{Gravito-electromagnetism}

Another surprising correspondence occurs when looking at the electric and magnetic fields for an observer co-moving with the dust source. These fields are
\begin{equation}
 \vec{E} = 0, \quad \vec{B} = \sqrt{2} \omega \hat{z}. \label{EMfields}
\end{equation}
The correspondence is with the field of gravito-electromagnetism \cite{PhysRevD.78.024021,Costa2014}, which concerns itself with analogies in tidal tensors between electromagnetism and gravity. This program has a very similar scope to the classical double copy: it draws analogies including but not limited to correspondences between linearized gravity and electromagnetism, as well as the structure of the Weyl and Maxwell tensors. Concretely one such analogy (following \cite{PhysRevD.78.024021}) analyzes tidal forces, which describe the relative acceleration of nearby particles with identical 4-velocity, which in electromagnetism and gravity take the respective forms:
\begin{equation}
    \frac{q}{m}\nabla_{\beta}(\tensor{F}{^\alpha_\mu} )u^\mu \delta x^\beta \hspace{5mm} \longleftrightarrow \hspace{5mm} -\tensor{R}{_{\mu \beta \nu}^\alpha} u^\mu u^\nu \delta x^\beta
\end{equation}
The Riemann tensor can be divided into ``electric" ($\mathbb{E}_{\alpha \beta}$) and ``magnetic" ($\mathbb{B}_{\alpha \beta}$) parts called gravitational tidal tensors, that follow analogous ``electric" ($E_{\alpha \beta}$) and ``magnetic" ($B_{\alpha \beta}$) tidal tensors from electromagnetism defined as:
\begin{align*}
    \tensor{E}{^\alpha_\beta} \equiv \nabla_\beta (\tensor{F}{^\alpha_\mu} )u^\mu \hspace{5mm} &\longleftrightarrow \hspace{5mm} \tensor{\mathbb{E}}{^\alpha_\beta} \equiv \tensor{R}{_{\mu \beta \nu}^\alpha} u^\mu u^\nu\\
    \tensor{B}{^\alpha_\beta} \equiv \nabla_\beta (\star \tensor{F}{^\alpha_\mu} )u^\mu \hspace{5mm} &\longleftrightarrow \hspace{5mm} \tensor{\mathbb{H}}{^\alpha_\beta} \equiv \star \tensor{R}{_{\mu \beta \nu}^\alpha} u^\mu u^\nu \numberthis
\end{align*}
Furthermore Maxwell's equations can be written in terms of these electromagnetic tidal tensors, and analogous equations hold for the gravitational tidal tensors. It turns out that such analogies can be made precisely in the G\"odel metric \cite{PhysRevD.78.024021,Costa2014}. In particular, the electromagnetic fields for an observer comoving with the dust source found through the tidal force analogy are precisely \eqref{EMfields}.\footnote{Note a change in metric conventions from \eqref{EMfields} to \cite{PhysRevD.78.024021,Costa2014}, $\omega \to \sqrt{2}\omega$.} We can thus also interpret the single-copy fields in the same manner as \cite{PhysRevD.78.024021}; nearby dust particles in the G\"odel universe rotate about one another, just as all charged particles with a nonzero velocity perpendicular to the magnetic field will undergo Larmor orbits. To the authors' knowledge, this is the first example of a direct connection between the classical single copy and the gravitoelectromagnetism program.
%instance of these two programs of analogy between general relativity and electromagnetism being connected.

\subsection{The flat limit of G\"odel}\label{SCgodelFlat}

Since the G\"odel metric is not of Kerr--Schild form, it is not guaranteed that the flat space versions of Maxwell's equations will yield the same form, an argument discussed in Section \ref{flatback}. Consider the electromagnetic tensor \eqref{Fdd} and its source \eqref{genKScurrents} with arbitrary $\kappa$ and $\theta$. Even if one assumes that $\kappa$ can have dependence on $\omega$, the flat limit is achieved as follows:
\begin{equation}
\lim_{\omega \to 0} \nabla^{(0)}_\mu \frac{1}{\kappa \omega} F^{\mu \nu} = \nabla^{(0)}_\mu \mathcal{F}^{\mu \nu} = 
 \lim_{\omega \to 0} \frac{1}{\kappa \omega}J_{(e)}^\nu = \lim_{\omega \to 0} -\sqrt{2}\omega \cos(\theta) \delta^{\mu}_\tau = 0. \label{flatproc}
\end{equation}
So, in our flat limit procedure, we find that the flat limit of the curved-background G\"odel single copy satisfies vacuum equations. Therefore, the aforementioned unsourced zeroth copy has no inconsistent interpretation on this flat background; the zeroth copy satisfies vacuum equations just as the single copy does. 

Using the source prescription from Kerr--Schild double copies yields the electromagnetic tensor and its source (on the original curved background) as:
\begin{equation}
 F_{\mu \nu} = \begin{pmatrix}
 0 & 0 & 0 & 0\\
 0 & 0 & -\sqrt{2}\omega \rho & 0\\
 0 & \sqrt{2}\omega \rho & 0 & 0\\
 0 & 0 & 0 & 0
 \end{pmatrix}, \hspace{10mm}
  J_{\text{(e)}}^\alpha = \begin{pmatrix}
2 \omega^2\\0\\0\\0
\end{pmatrix}.
\end{equation}
The flat limit $\omega \to 0$ under the procedure \eqref{flatproc} yields the same form of electromagnetic tensor except now in a flat cylindrical Minkowski space, for which such a tensor describes a constant magnetic field $\vec{B} \sim \hat{z}$ for the observer $u^\mu = \delta^\mu_\tau$, and therefore satisfies vacuum equations. We see that our flat space equations therefore maintain the form of our electromagnetic tensor, but it is now unsourced. The former fact is a result of nice coordinates. If instead we chose the coordinates of \eqref{godelMetricCyl}, we find an electromagnetic tensor of the form
\begin{equation}\label{fjeflatgodel}
 F_{\mu \nu} = \begin{pmatrix}
 0 & 0 & 0 & 0\\
 0 & 0 & -\sqrt{2}\sinh(\omega r) & 0\\
 0 & \sqrt{2}\sinh(\omega r) & 0 & 0\\
 0 & 0 & 0 & 0
 \end{pmatrix}, \hspace{10mm}
  J_{\text{e}}^\alpha = \begin{pmatrix}
2 \omega^2\\0\\0\\0
\end{pmatrix},
\end{equation}
for which the flat space procedure yields the same constant magnetic field in cylindrical Minkowski space,
\begin{equation}
\lim_{\omega \to 0} \frac{1}{\omega} F_{\mu \nu} =\lim_{\omega \to 0} \frac{1}{\omega} \begin{pmatrix}
 0 & 0 & 0 & 0\\
 0 & 0 & -\sqrt{2}\sinh(\omega r) & 0\\
 0 & \sqrt{2}\sinh(\omega r) & 0 & 0\\
 0 & 0 & 0 & 0
 \end{pmatrix} = \begin{pmatrix}
 0 & 0 & 0 & 0\\
 0 & 0 & -\sqrt{2} r & 0\\
 0 & \sqrt{2} r & 0 & 0\\
 0 & 0 & 0 & 0
 \end{pmatrix},
\end{equation}
but with the radial coordinate identified as $r$ rather than $\rho$. We stress the importance of a curved space covariant derivative in this process; since $\nabla_\mu$ introduces factors of $\omega$ into the current $J^\mu$, a flat limit will not include these higher order terms in $\omega$, and hence the current will vanish in the flat limit.

\subsection{Introducing coordinate dependence}\label{godelFuncIntro}

So far, we have only considered a constant electromagnetic spinor for the single copy. However, in principle it may be a complex function of the coordinates. We may ask: could other electromagnetic spinors admit the Kerr--Schild prescription's sources? Allowing for this possibility while still maintaining the symmetries of the spacetime within the single-copy will force upon us the curved-background Maxwell spinor \eqref{KSfAB}. To begin, we write the electromagnetic spinor and scalar in the Weyl double copy as
\begin{equation}
 f_{AB} = i \omega \Omega(x^\mu) o_{(A}\iota_{B)}, \qquad S = \Omega(x^\mu)^2. \label{genfuncdep}
\end{equation}
Consider the curved background case, and let $\Omega(x^\mu) = \Omega_R(x^\mu) + i \Omega_I(x^\mu)$. The electromagnetic field strength is
\begin{equation}
 F_{\mu \nu} =\omega  \begin{pmatrix}
0 & 0 & 0 & - \Omega_I\\
0 & 0 & \rho \, \Omega_R & 0\\
0 & - \rho \, \Omega_R & 0 & -\frac{1}{\sqrt{2}}\omega \rho^2  \Omega_I\\
 \Omega_I & 0 & \frac{1}{\sqrt{2}}\omega \rho^2  \Omega_I & 0
\end{pmatrix}, \label{Fddgen}
\end{equation}
writing $\Omega_R = \Omega_R(x^\mu)$ and $\Omega_I = \Omega_I(x^\mu)$ for notational convenience. The associated currents are:
\begin{align}
 J^\mu_{\text{e}} &= \frac{\omega}{\sqrt{2}} \begin{pmatrix}
 -\sqrt{2}\partial_z \Omega_I -2 \omega \Omega_R - \omega \rho \, \partial_\rho \Omega_R\\
 -\sqrt{2}\rho^{-1} \partial_\phi \Omega_R + \omega \rho \, \partial_\tau \Omega_R\\
 \sqrt{2}\rho^{-1} \partial_\rho \Omega_R\\
 \sqrt{2}\partial_\tau \Omega_I
 \end{pmatrix}, \label{Jegen}\\
 J^\mu_{\text{m}} &= \frac{\omega}{\sqrt{2}} \begin{pmatrix}
 \sqrt{2}\partial_z \Omega_R -2 \omega \Omega_I - \omega \rho \, \partial_\rho \Omega_I\\
 -\sqrt{2}\rho^{-1} \partial_\phi \Omega_I + \omega \rho \, \partial_\tau \Omega_I\\
 \sqrt{2}\rho^{-1} \partial_\rho \Omega_I\\
 -\sqrt{2}\partial_\tau \Omega_R
 \end{pmatrix}.\label{Jmgen}
\end{align}
If we assume that the symmetries of the spacetime (axisymmetry) are obeyed in the single copy, then $\Omega(x^\mu) = \Omega(\rho)$ and only partial derivatives involving $\rho$ will remain. Upon applying the Kerr--Schild source prescription, we find that
\begin{equation}
 J^\varphi_{\text{e}} \sim \partial_\rho \Omega_R = 0,
\end{equation}
and so $\Omega_R$ must be a constant. This implies via the Hodge dual of the sourced Maxwell equations that $\Omega_I$ is also a constant, and so the electric current is
\begin{equation}
 J_{\text{e}}^\alpha =  \begin{pmatrix}
-\omega^2 \sqrt{2} \Omega_R\\
0\\
0\\
0
\end{pmatrix} = \begin{pmatrix}
2\omega^2\\
0\\
0\\
0
\end{pmatrix},
\end{equation}
To ensure consistency with the Kerr--Schild source \eqref{KSfAB}, which in turn had given the current \eqref{fjeflatgodel}, we set $\Omega_R = -\sqrt{2}$. That is to say, there does not exist a more general solution adhering to the symmetries of the spacetime that results in the Kerr--Schild prescription sources. 

\subsection{The electromagnetic interpretation of G\"odel and some properties}

We now come to the electromagnetic interpretation of the G\"odel metric through our single copy correspondence outlined in the previous sections, which has an elegant interpretation in both the original curved and flat limit background. On the original curved space, the electric current (setting magnetic current to zero) is a constant charge density. We reiterate the ``everywhere-rotating" nature of the background on which these sources live, meaning that along the worldline of an observer co-moving with a dust particle, any other dust particle has a constant vorticity vector. Therefore, a uniform charge density effectively rotates about such an observer, resulting in a constant magnetic field for the observer.

The other aspects of the electromagnetic field are most easily gleaned from imposing the flat limit case described in Section \ref{flatback}. This is because this procedure removes the curvature, and hence the gravitational effects present in the background spacetime, isolating the purely electromagnetic effects. In the case of G\"odel, we found that this yields a constant magnetic field in the $z$-direction. Recall the four defining features of the G\"odel metric which we seek to form a single copy interpretation for: (I) it is stationary and axisymmetric; (II) it is a homogeneous space; (III) it represents an ``everywhere rotating" dust source; and, (IV) along any chosen origin, a natural circular null curve arises, separating a region of closed spacelike curves and closed timelike curves.

(I) is a consequence of the Weyl double copy procedure that we follow; since the metric is independent of the coordinates $\{\tau, \phi, z\}$, so too will be the electromagnetic spinor constructed from $\Psi_2$. Note that in the general case where we introduce new coordinate dependence into the Weyl double copy, this condition must be enforced, rather than emergent. (II) is represented through the single copy being an everywhere homogeneous magnetic field in flat space. (III) was discussed in the paragraph following \eqref{EMfields}, and we repeat its main points here. In a constant magnetic field, all charged particles with any perpendicular velocity to the field will undergo a rotation, aka ``Larmor orbits." This is how ``everywhere rotation" manifests itself in the single copy, and we emphasize that this interpretation was first made in \cite{PhysRevD.78.024021}.

(IV) however does not have a natural interpretation in our single copy, and we make a brief remark as to why our flat limit has no remnant of closed time-like curves. Recall that the radius at which closed timelike curves exist in the original curved spacetime is beyond the radius $\rho_{\text{CNC}} \sim \omega^{-1}$ defined in \eqref{rhocnc}, and so taking the flat limit $\omega \to 0$ sends these closed timelike curves to infinity. Therefore, it is not possible to disentangle the gravitational effects from the electromagnetic ones.

\subsection{Bootstrapping the fields on flat space through symmetry}\label{flatspacesec}

If one hopes to hold on to a sourced Weyl double copy on a \textit{flat} background, we cannot simply use the Kerr--Schild source prescription despite it yielding a nice interpretation on the curved background (see Sections \ref{SCgodelFlat} and \ref{godelFuncIntro}), since the sources inevitably vanish or yield results which do not respect the symmetries of the spacetime. To explicitly see this, consider the flat limit of our electromagnetic sources with general coordinate dependence \eqref{Jegen}--\eqref{Jmgen},
\begin{equation}
 \lim_{\omega \to 0} \frac{1}{\omega} J^\mu_{\text{e}} = \frac{1}{\sqrt{2}} \begin{pmatrix}
 -\sqrt{2}\partial_z \Omega_I \\
 -\sqrt{2}\rho^{-1} \partial_\phi \Omega_R \\
 \sqrt{2}\rho^{-1} \partial_\rho \Omega_R\\
 \sqrt{2}\partial_\tau \Omega_I
 \end{pmatrix},\ \ \ \ 
\lim_{\omega \to 0} \frac{1}{\omega} J^\mu_{\text{m}} = \frac{1}{\sqrt{2}} \begin{pmatrix}
 \sqrt{2}\partial_z \Omega_R \\
 -\sqrt{2}\rho^{-1} \partial_\phi \Omega_I\\
 \sqrt{2}\rho^{-1} \partial_\rho \Omega_I\\
 -\sqrt{2}\partial_\tau \Omega_R
 \end{pmatrix},
\end{equation}
for which a dependence on $z$ within $\Omega_I$ must be introduced to allow a source of the form $J^\mu \sim \delta^\mu_\tau$. Instead, let us leverage the symmetries of the spacetime to see which sources are allowed. Let us start by bootstrapping the single copy with a flat background, with the general functional dependence \eqref{genfuncdep}. One can consider the limit $\omega \to 0$ for the vierbein given by (\ref{cplxg1}-\ref{cplxg3}), or by considering the flat limit of the general electromagnetic field \eqref{Fddgen}. Either way, we get:
\begin{equation}
F_{\alpha\beta} = \omega\begin{pmatrix}
0 & 0 & 0 & -\Omega_{\text{I}}\\
0 & 0 & \rho\Omega_{\text{R}} & 0\\
0 & -\rho\Omega_{\text{R}} & 0 & 0\\
\Omega_{\text{I}} & 0 & 0 & 0
\end{pmatrix},\ \ \ \ (\alpha,\beta = \tau,\rho,\phi,z).
\end{equation}
We can also compute the Hodge dual $\tilde{F}$ (with the convention $\epsilon_{tr\phi z} = \rho$):
\begin{equation}
\tilde{F}_{\alpha\beta} \equiv \frac{1}{2}\tensor{\epsilon}{_\alpha_\beta_\gamma_\delta}F^{\gamma\delta} = \omega\begin{pmatrix}
0 & 0 & 0 & \Omega_{\text{R}}\\
0 & 0 & \rho\Omega_{\text{I}} & 0\\
0 & -\rho\Omega_{\text{I}} & 0 & 0\\
-\Omega_{\text{R}} & 0 & 0 & 0
\end{pmatrix}.
\end{equation}
Now, we plug the expressions for the electromagnetic tensor, its Hodge dual, and $S$ into the equations \eqref{maxwell} with a flat background, so as to relate $\Omega$ to the putative sources. We will also introduce a zeroth copy $S = \Omega^2 = J_{\text{e}}^\tau + i J_{\text{m}}^\tau$ to make contact with the Kerr--Schild procedure. All together:
\begin{equation}
J_{\text{e}}^\alpha = \frac{\omega}{\rho} \begin{pmatrix}
-\rho\,\partial_z \Omega_{\text{I}}\\
-\partial_\phi \Omega_{\text{R}}\\
\partial_\rho \Omega_{\text{R}}\\
\rho\,\partial_t \Omega_{\text{I}}
\end{pmatrix},\ \ J_{\text{m}}^{\alpha} = \frac{\omega}{\rho}\begin{pmatrix}
\rho\,\partial_z \Omega_{\text{R}}\\
-\partial_\phi \Omega_{\text{I}}\\
\partial_\rho \Omega_{\text{I}}\\
-\rho\,\partial_t \Omega_{\text{R}}
\end{pmatrix},\ \ J_{\text{e}}^\tau = \square^{(0)}\left[\Omega_{\text{R}}^2 - \Omega_{\text{I}}^2\right],\ \ J_{\text{m}}^\tau = \square^{(0)}\left[2\Omega_{\text{R}}\Omega_{\text{I}}\right].\label{currentwaveeqngen}
\end{equation}
This is the stage where we impose both time-independence and axisymmetry. These are both symmetries of the G\"odel metric itself, and so we also assume that they manifest in the single-copy gauge field. Mathematically, this amounts to
\begin{equation}
\Omega_{\text{R}} = \Omega_{\text{R}}(\rho),\ \ \ \ \Omega_{\text{I}} = \Omega_{\text{I}}(\rho).
\end{equation}
The currents and wave equation in \eqref{currentwaveeqngen} become
\begin{equation}
\begin{split}
&J_{\text{e}}^{\alpha} = \frac{\omega}{\rho}\begin{pmatrix}
0\\
0\\
\Omega_{\text{R}}'(\rho)\\
0
\end{pmatrix},\ \ \ \ 
J_{\text{m}}^{\alpha} = \frac{\omega}{\rho}\begin{pmatrix}
0\\
0\\
\Omega_{\text{I}}'(\rho)\\
0
\end{pmatrix},\\
&\frac{1}{\rho}\partial_\rho\left[\rho \partial_\rho \left(\Omega_{\text{R}}^2 - \Omega_{\text{I}}^2\right)\right] = 0,\ \ \ \ \frac{1}{\rho}\partial_\rho\left[\rho \partial_\rho \left(2\Omega_{\text{R}}\Omega_{\text{I}}\right)\right] = 0.
\end{split}
\end{equation}
We have no source for the zeroth-copy field. This is because we are assuming that it should match the charge density in the source of the single-copy fields, but with axisymmetry both the electric and magnetic charge densities (respectively $J_{\text{e}}^t$ and $J_{\text{m}}^t$) are zero. With this in mind, the components of the wave equation yields
\begin{equation}
\Omega_{\text{R}}^2 - \Omega_{\text{I}}^2 = A\log\left(\frac{\rho}{B}\right),\ \ \ \ 2\Omega_{\text{R}}\Omega_{\text{I}} = C\log\left(\frac{\rho}{D}\right),
\end{equation}
where $\{A,B,C,D\}$ are real constants. We solve these equations for the components of $\Omega$ to write
\begin{equation}
\begin{split}
\Omega_{\text{R}}(\rho) &= \pm\sqrt{\frac{1}{2} \left[A\log (\frac{\rho}{B}) + \sqrt{\left( A\log(\frac{\rho}{B}) \right)^2 + \left( C\log(\frac{\rho}{D}) \right)^2} \right]},\\
\Omega_{\text{I}}(\rho) &= \pm\sqrt{\frac{1}{2} \left[-A\log (\frac{\rho}{B}) + \sqrt{\left( A\log(\frac{\rho}{B}) \right)^2 + \left( C\log(\frac{\rho}{D}) \right)^2} \right]}.
\end{split}\label{genomegas}
\end{equation}
The currents in the $\phi$ direction are essentially derivatives of these functions, and the electric and magnetic fields in the $z$ direction for a timelike observer $u^\mu = \delta^\mu_\tau$:
\begin{equation}
E_z = -\omega \Omega_{\text{I}}(\rho),\ \ \ \ B_z = \omega \Omega_{\text{R}}(\rho).
\end{equation}

\paragraph{Physical interpretation} That this single copy has such an odd form might be attributed to the odd causal properties of G\"odel. However if we evaluate this single copy against the defining properties of the spacetime, other than its symmetries, we have no natural interpretation. The fields are clearly not homogeneous, there is a preferred origin, and while the integration constants $B,D$ do define some radial scale, it is unclear how these relate in any way to the scale set by the spacetime. If we are to hope for some correspondence, we therefore seek for a relation between the closed timelike curves of G\"odel and the preferred radial scale. A possible (and physically reasonable) road to simplification is for the magnetic currents to vanish, which is attainable via $\Omega_I = 0$ and
\begin{equation}
\Omega_{\text{R}} = \begin{cases}
\pm\sqrt{A\log\left(\dfrac{\rho}{\rho_*}\right)}&\text{if}\ \ \rho > \rho_*,\vspace{0.2cm}\\
\pm \sqrt{-A\log\left(\dfrac{\rho}{\rho_*}\right)}&\text{if}\ \ \rho < \rho_*.
\end{cases}\label{omegarnomag}
\end{equation}
For the $+$ choice in \eqref{omegarnomag}, we get the following Lorentz force equations for a particle with charge $q$ in flat space experiencing the magnetic field $B_z=\omega\Omega_{\text{R}}(\rho)$:
\begin{align}
&\frac{d u^t}{d \lambda} = \frac{du^z}{d\lambda} = 0,\label{geo1nomag}\\
&\frac{d u^\rho}{d \lambda} - \rho u^\phi\left(\frac{q}{m} \omega\sqrt{\pm A \log\left(\frac{\rho}{\rho_*} \right)} + u^\phi\right) = 0,\label{geo2nomag}\\
&\frac{d u^\phi}{d \lambda} + \frac{u^\rho}{\rho}\left(\frac{q}{m} \omega\sqrt{\pm A \log\left(\frac{\rho}{\rho_*} \right)} + 2u^\phi\right) = 0,\label{geo3nomag}
\end{align}
where here $+$ denotes the $\rho>\rho_*$ branch and $-$ denotes the $\rho<\rho_*$ branch. In contrast, instead choosing the $-$ sign in \eqref{omegarnomag} introduces additional $\pm$ symbols on any square-root terms. We can realize closed circular orbits with $u^z = 0$ for the case of no magnetic current, since the geodesic equations here imply constant $u^t$ and $u^z$. From the equations \eqref{geo2nomag} and \eqref{geo3nomag}, if $u^\rho = 0$ then
\begin{equation}
u^\phi = -\frac{q}{m} \omega\sqrt{\pm A \log\left(\frac{\rho_{\text{circ}}}{\rho_*} \right)} \implies \rho_{\text{circ}} = \rho_* \exp \left[\pm \frac{1}{A} \left( \frac{m u^\phi}{\omega q} \right)^2 \right],\label{rcircformula}
\end{equation}
which evidently has no correspondence with the form \eqref{cncEq} or \eqref{rhocnc}.

\section{Conclusions and outlook}\label{sec:concls}

In this work, we have shown that the current examples within the classical double copy do not distinguish between the original curved background of the spacetime or an appropriate flat background for the single-copy fields. That such an ambiguity exists is due to the Kerr--Schild (or linearized) form in which all analyzed spacetimes have been expressible. We have presented a framework through which such appropriate flat background Maxwell fields may be constructed, showing that when stepping away from a Kerr--Schild form, equivalent Maxwell equations between the originally curved space and appropriate flat background are not guaranteed to hold.

Since the flat background is achieved through a limiting procedure (not recoverable vice-versa), we assert that single-copy fields generally should be constructed on the original curved background. As noted in the Introduction, this is not the same statement as creating an electromagnetic field that consistently sources the original curved spacetime, but rather a statement about curvature quantities and Einstein's equations in certain situations being capable of being rewritten as electromagnetic fields satisfying Maxwell's equations. A consequence of this step is dropping the previously defined zeroth copy correspondence, $\Box\psi\sim R^0_{ \ 0}$, since on flat space the wave equation sourced by electric charge density is an immediate consequence of the flat space single-copy equations, whereas the zeroth copy correspondence is not generally true in curved space (even for Kerr--Schild metrics).

The G\"odel universe, which satisfies sourced Maxwell equations on the curved background but vacuum equations in the flat limit, is a simple non-Kerr--Schild spacetime that demonstrates the lack of equivalence between single copies on these backgrounds. During this investigation, we found that a natural source emerges, specifically related to the mixed index Ricci tensor in the same manner as the Kerr--Schild double copy, despite the metric not having a (geodesic) Kerr--Schild form. This correspondence revealed a simple and satisfying single copy for the G\"odel spacetime, namely a constant magnetic field. This constant magnetic field shares the same defining properties as the spacetime: the symmetries, homogeneity, and the presence of Larmor orbits of particles at every point (representative of the ``everywhere rotating" feature). In a surprising equivalence, we found that this single copy directly matches the electromagnetic field from tidal force gravitoelectromagnetic analogies \cite{PhysRevD.78.024021,Costa2014}; this appears to be the first explicit connection made between these two unique GR-EM correspondences.

On the other hand, we also investigated another single-copy construction, forcing a flat space sourced single copy to exist by introducing new functional dependencies and enforcing the symmetries of the spacetime. We found that such a sourced flat-space bootstrap approach not only could not be sourced via the same prescription as Kerr--Schild metrics, but also failed to share the same defining properties of the spacetime, further strengthening the claim that single-copy fields should be constructed on the original curved background.

We emphasize that this paper is only a small step towards a more fundamental understanding of the classical double copy. Twistors may play a fundamental role in such an understanding \cite{Chacon:2021wbr,Chacon:2021lox,Luna:2022dxo}, and with our analysis on backgrounds perhaps curved space twistors may elucidate an answer. However, even at the level of spinorial general relativity, which was taken as foundational in this work, much is left to be said on the conditions under which the classical double copy can exist. For instance, the stationary condition is a seemingly necessary requirement for the Kerr--Schild double copy to exist with sources, and that our electromagnetic source had a similar form was also predicated on the fact that G\"odel is stationary. While it has been shown that in vacuum \cite{Easson:2023dbk} one may circumvent the stationary condition, a more fundamental understanding of how this source prescription arises in general spacetimes is desireable. Whether or not Killing's equation and the geodesic condition are fundamental to single-copies, such that single-copies are simply formulated from just these principles, must be investigated further. Furthermore, that same work demonstrated Type II Kerr--Schild double copies are possible, thus the Weyl double copy relations for Type II spacetimes will be interesting to explore, as was initiated in \cite{Keeler:2024bdt} within the fluid-gravity correspondence. It will additionally be interesting to explore if the classical double copy is restricted to algebraically special spacetimes, and hence a consequence of having aligned principal null directions (perhaps in conjunction with the other assumptions). We hope this work initiates the pursuit of a more general understanding of the classical double copy beyond Kerr--Schild.

\vspace{5mm}

\noindent\textbf{Acknowledgments} The authors thank Gilly Elor, Gabriel Herczeg, Ricardo Monteiro, Donal O'Connell, Andrew Svesko, and Aaron Zimmerman for useful comments on the draft. B.K. was supported by NSF grant PHY–2210562. T.M. was supported by Simons Foundation, Award 896696 while this work was in progress. S.S. was supported by NSF grants PHY-2112725 and PHY-2210562, DOE grant DE-SC001010, and the Federico and Elvia Faggin Foundation.

\bibliographystyle{jhep}
\bibliography{refs.bib}

\end{document}